\pdfoutput=1
\documentclass[11pt,twoside,a4paper,cmspaper,final,collab]{cms-tdr}
\def\svnVersion{b84376a}\def\svnDate{2025/01/24}\def\cmsCernNoTag{CERN-EP-2023-258}\def\cmsCernDate{\today}\def\cmsMessage{Published in Physical Review D as \href{http://dx.doi.org/10.1103/PhysRevD.111.012009}{\doi{10.1103/PhysRevD.111.012009}.}}
\begin{document}\cmsNoteHeader{TOP-22-005}

\newcommand{\eee}{\ensuremath{\Pe\Pe\Pe}\xspace}
\newcommand{\mmm}{\ensuremath{\PGm\PGm\PGm}\xspace}
\newcommand{\eml}{\ensuremath{\Pe\PGm\Pell}\xspace}
\providecommand{\PY}{{\HepParticle{Y}{}{}}\Xspace}
\newlength\cmsTabSkip
\setlength{\cmsTabSkip}{0.8ex}
\newlength{\ruleht}
\setlength{\ruleht}{\baselineskip}
\cmsNoteHeader{TOP-22-005}

\title{Search for charged-lepton flavor violation in the production and decay of top quarks using trilepton final states in proton-proton collisions at \texorpdfstring{$\sqrt{s} = 13\TeV$}{sqrt(s) = 13 TeV}}

\date{\today}

\abstract{
A search is performed for charged-lepton flavor violating processes in top quark (\PQt) production and decay. The data were collected by the CMS experiment from proton-proton collisions at a center-of-mass energy of 13\TeV and correspond to an integrated luminosity of 138\fbinv. The selected events are required to contain one opposite-sign electron-muon pair, a third charged lepton (electron or muon), and at least one jet of which no more than one is associated with a bottom quark. Boosted decision trees are used to distinguish signal from background, exploiting differences in the kinematics of the final states particles. The data are consistent with the standard model expectation. Upper limits at 95\% confidence level are placed in the context of effective field theory on the Wilson coefficients, which range between 0.024--0.424$\TeV^{-2}$ depending on the flavor of the associated light quark and the Lorentz structure of the interaction. These limits are converted to upper limits on branching fractions involving up (charm) quarks, $\PQt\to\Pe\PGm\PQu$ ($\PQt\to\Pe\PGm\PQc$), of $0.032 (0.498) \times 10^{-6}$, $0.022 (0.369) \times 10^{-6}$, and $0.012 (0.216) \times 10^{-6}$ for tensor-like, vector-like, and scalar-like interactions, respectively.}

\hypersetup{%
pdfauthor={CMS Collaboration},%
pdftitle={Search for charged-lepton flavor violation in the production and decay of top quarks using trilepton final states in proton-proton collisions at sqrt(s) = 13 TeV},%
pdfsubject={CMS},%
pdfkeywords={CMS, TOP, LFV, EFT}} %

\maketitle

\section{Introduction}
\label{sec:introduction}

Lepton flavor is conserved in the standard model (SM) with massless neutrinos. However, the observation of neutrino oscillations~\cite{Super-Kamiokande:1998kpq,SNO:2002tuh} confirms the existence and the mixing of massive neutrinos, and it also indicates that lepton flavor violation (LFV) is expected to occur in the charged-lepton sector. 
The CERN LHC can provide sensitivity~\cite{Davidson:2012wn} to charged-LFV (CLFV) searches, including in two- or three-body decays of heavy particles, ${\PX}\to\Pell\Pell^{\prime}(\PY)$, and in heavy-particle production, $\Pp\Pp \to \Pell\Pell^{\prime}\PX$. Here, \PX refers to a heavy SM particle such as a top quark ($\PQt$) or a Higgs, \PW, or \PZ boson, $\Pell(\Pell^{\prime}$) are charged leptons with different flavor and opposite charge, \PY denotes an additional generic SM particle, and \Pp stands for proton. 
For CLFV processes involving the heaviest of all elementary particles, the top quark, competitive sensitivity is predicted at the LHC compared to previous bounds on such interactions~\cite{Davidson}. 
Recent flavor anomalies in decays of \PB mesons, including the recent potential lepton flavor universality violation reported by the LHCb experiment~\cite{LHCb:2023zxo}, has prompted renewed experimental interest in CLFV searches. Moreover, some models that accommodate these flavor anomalies also suggest that CLFV involving a top quark is within the reach of the LHC sensitivity~\cite{Kim:2018oih}. This paper presents a search for CLFV processes in top quark production and decay. The search utilizes the $\Pp\Pp$ collision data collected by the CMS experiment at the LHC in 2016--2018 at a center-of-mass energy of 13\TeV, corresponding to an integrated luminosity of 138\fbinv. 

\begin{figure*}[tbhp]
 \centering
  \includegraphics[width=0.31\textwidth]{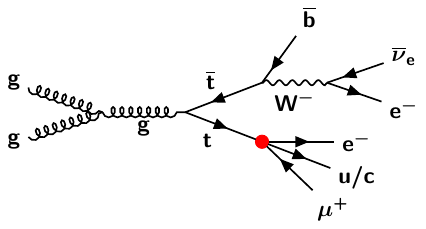}
  \includegraphics[width=0.33\textwidth]{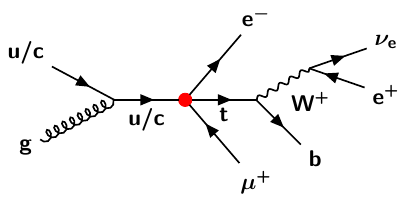}
  \includegraphics[width=0.31\textwidth]{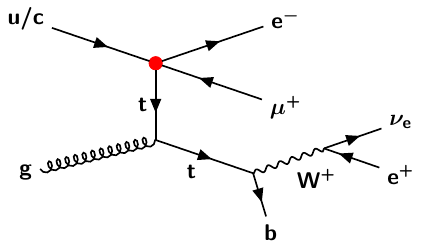}
\caption{Representative Feynman diagrams for the signal processes that are targeted by this analysis. Both top quark decay (left) and production (middle and right) CLFV processes are shown. The CLFV interaction vertex is shown as a solid red circle to indicate that it is not allowed in the SM. The charge conjugate diagrams also contribute.}
\label{fig:Feynman_Diagram}
\end{figure*}

The signal processes considered are shown with representative Feynman diagrams in Fig.~\ref{fig:Feynman_Diagram}. 
Final states with exactly three charged leptons, either electrons ($\Pe$) or muons ($\PGm$), 
are considered. One of the three leptons is presumed to originate from the leptonic decay of the SM top quark and the other two from CLFV interactions. While the signal models considered do not include tau leptons ($\Pgt$), tau leptons from SM top quark decays that subsequently decay into an electron or a muon do contribute. 
In addition, the selected events are required to contain at least one jet of which no more than one is consistent with originating from a bottom ($\Pb$) quark. 

The CMS Collaboration has previously searched for CLFV in the top quark sector utilizing final states with two oppositely charged leptons~\cite{TOP-19-006}. This paper, which requires three leptons in the final state, follows a similar strategy to parametrize CLFV signals with dimension-6 effective field theory (EFT) operators, 
\begin{equation}
\mathcal{L}=\mathcal{L}_{\text{SM}}^{(4)} + \frac{1}{\Lambda^2}\sum_{a}C_{a}^{(6)}O_{a}^{(6)}+O\left( \frac{1}{\Lambda^4}\right),
\end{equation}  
where $\mathcal{L}_{\text{SM}}^{(4)}$ is the renormalizable Lagrangian of the SM. The $O_{a}^{(6)}$ denote dimension-6 nonrenormalizable operators, and $C_{a}^{(6)}$ are the corresponding Wilson coefficients. The dimension-6 terms are suppressed by the square of a mass scale $\Lambda$ where new physics is presumed to emerge. Assuming the CLFV coupling involves exactly one electron, one muon, one top quark, and one up (\PQu)/charm (\PQc) quark, the complete list of dimension-6 operators \cite{Grzadkowski} can be shortened. We summarize the relevant operators in Table~\ref{tab:dimension6}.

\begin{table}[tbh]
\centering
\topcaption{Summary of relevant dimension-6 operators considered in this analysis. Here, $\varepsilon$ is the two-dimensional Levi-Civita symbol, $\gamma^\mu$ the Dirac gamma matrices, and $\sigma^{\mu\nu}=\frac{i}{2}[\gamma^\mu,\gamma^\nu]$. The $\PQq$ and $\Pl$ denote left-handed doublets for leptons and quarks, respectively, whereas $\PQu$ and $\Pe$ denote right-handed singlets for quarks and leptons, respectively. The indices $i$ and $j$ are lepton flavor indices that run from 1 to 2 with $i \neq j$; $k$ and $l$ are quark flavor indices with the condition that one of them is 3 and the other one is 1 or 2. The four vector-like operators are merged in this analysis because the final-state particles produced by these operators have very similar kinematics.}
\begin{scotch}{cccl}
Lorentz structure & \multicolumn{3}{c}{Operator}\\
\hline
\rule[-0.2\ruleht]{0pt}{1.15\ruleht}\multirow{4}{*}{Vector} & $O_{\Pl\PQq}^{(1)ijkl}$ &=& $(\bar{\Pl}_i\gamma^\mu \Pl_j)
   ({\PAQq}_k\gamma_\mu \PQq_l)$
   \\
  & $O_{\Pl\PQu}^{ijkl}$ &=& $(\bar{\Pl}_i\gamma^\mu \Pl_j)
   ({\PAQu}_k\gamma_\mu \PQu_l)$
   \\  
  & $O_{\Pe\Pq}^{ijkl}$ &=& $(\bar{\Pe}_i\gamma^\mu \Pe_j)
   ({\PAQq}_k\gamma_\mu \PQq_l)$
   \\ 
  & $O_{\Pe\PQu}^{ijkl}$ &=& $(\bar{\Pe}_i\gamma^\mu \Pe_j)
   ({\PAQu}_k\gamma_\mu \PQu_l)$
   \\
\hline
\rule[-0.2\ruleht]{0pt}{1.15\ruleht}\multirow{1}{*}{Scalar}  & $O_{\Pl\Pe\PQq\PQu}^{(1)ijkl}$ &=& $(\bar \Pl_i \Pe_j)\;\varepsilon\;
   (\bar \PQq_k u_l)$
   \\ 
   \hline
   \rule[-0.2\ruleht]{0pt}{1.15\ruleht}\multirow{1}{*}{Tensor} & $O_{\Pl\Pe\PQq\PQu}^{(3)ijkl}$ &=& $(\bar \Pl_i \sigma^{\mu\nu} \Pe_j)\;\varepsilon\;
   ({\PAQq}_k \sigma_{\mu\nu} \PQu_l)$
\end{scotch}
\label{tab:dimension6}
\end{table}

The background processes that enter the final selection are categorized into two groups: prompt and nonprompt backgrounds. Prompt backgrounds include SM processes that produce at least three leptons via decays of electroweak gauge bosons. Processes that fail to meet this criterion are categorized as nonprompt backgrounds. Prompt backgrounds in this analysis are modeled with Monte Carlo (MC) simulation while the nonprompt backgrounds are modeled using control samples in data. 
 
To distinguish CLFV signal and background events, machine learning techniques, specifically boosted decision trees (BDTs), are employed. The full BDT discriminant distributions are used to construct a binned likelihood function, which is used to interpret the observed data in terms of limits on Wilson coefficients and branching fractions associated to various signal processes.  

The paper is organized as follows. Section~\ref{sec:CMS} gives a brief overview of the experimental apparatus and Section~\ref{sec:samples} describes the data and simulated samples used for the analysis. The reconstruction of physics objects is presented in Section~\ref{sec:objects} followed by the event selection in Section~\ref{sec:selection}. Descriptions of the background estimation and signal extraction are given in Sections~\ref{sec:background} and~\ref{sec:BDT}, respectively. The systematic uncertainties that contribute to this analysis are outlined in Section~\ref{sec:systematics}. Finally, the limit-setting procedure and results are discussed in Section~\ref{sec:results}, and a summary is given in Section~\ref{sec:summary}.

\section{The CMS detector}
\label{sec:CMS}

The central feature of the CMS apparatus is a superconducting solenoid of 6\unit{m} internal diameter, providing a magnetic field of 3.8\unit{T}. Within the solenoid volume are a silicon pixel and strip tracker, a lead tungstate crystal electromagnetic calorimeter (ECAL), and a brass and scintillator hadron calorimeter (HCAL), each composed of a barrel and two endcap sections. Forward calorimeters extend the pseudorapidity ($\eta$) coverage provided by the barrel and endcap detectors. Muons are measured in gas-ionization detectors embedded in the steel flux-return yoke outside the solenoid. A more detailed description of the CMS detector, together with a definition of the coordinate system used and the relevant kinematic variables, can be found in Ref.~\cite{CMS:2008xjf}.

Events of interest are selected using a two-tiered trigger system. The first level (L1), composed of custom hardware processors, uses information from the calorimeters and muon detectors to select events at a rate of around 100\unit{kHz} within a fixed latency of about 4\mus~\cite{CMS:2020cmk}. The second level, known as the high-level trigger, consists of a farm of processors running a version of the full event reconstruction software optimized for fast processing, and reduces the event rate to around 1\unit{kHz} before data storage~\cite{CMS:2016ngn}.

\section{Data and simulated samples}
\label{sec:samples}

The $\Pp\Pp$ collision data used in this search were recorded with a large number of triggers that require the presence of one, two, or three electrons or muons to ensure a high trigger efficiency. Only data collected during periods when the CMS detector was fully operational are included. The resulting integrated luminosity for the 2016--2018 data set is 138\fbinv.

To model the signal and to estimate the contributions from some of the background processes, MC simulations are used. 
Independent samples are generated for the years 2016, 2017, and 2018 to match the different data-taking conditions.
The signal events are generated at leading order using \MGvATNLO 2.6.5~\cite{Alwall_2014}, which makes use of the external model \textsc{SmeftFR}~\cite{Dedes:2019uzs} to incorporate EFT effects. The events generated with \textsc{SmeftFR} are reweighted event-by-event to match the predictions from \textsc{SMEFTsim}~\cite{Brivio:2017btx}.

Utilizing the strategy described in Ref.~\cite{TOP-19-006}, we can describe the potential CLFV couplings with exactly six independent Wilson coefficients: $C_{\Pe\PGm\PQt\PQu}^{\text{tensor}}$, $C_{\Pe\PGm\PQt\PQu}^{\text{vector}}$, $C_{\Pe\PGm\PQt\PQu}^{\text{scalar}}$, $C_{\Pe\PGm\PQt\PQc}^{\text{tensor}}$, $C_{\Pe\PGm\PQt\PQc}^{\text{vector}}$, $C_{\Pe\PGm\PQt\PQc}^{\text{scalar}}$, where the subscript describes the CLFV interaction vertex and the superscript indicates the Lorentz structure of the operators. The signal samples are produced for each of these six couplings, separately for production and decay signal modes, with an arbitrary choice of $C_a=1$ and $\Lambda=1\TeV$ for all Wilson coefficients during the event generation. Theoretical cross sections for the different signal processes are shown in Table~\ref{tab:XS}. The quoted uncertainties are related to the choice of parton distribution function (PDF) and QCD renormalization and factorization scales. The top quark production signal cross sections are determined with \textsc{SMEFTsim}~\cite{Brivio:2017btx}, while the top quark decay signal cross sections are obtained by: 
\begin{equation}
\sigma_{\text{CLFV}}=4\,\sigma(\Pp\Pp\to\ttbar)_{\text{SM}}\,\mathcal{B}(\PQt \to \Pe\PGm\PQq),
\end{equation}
where $\sigma(\Pp\Pp\to\ttbar)_{\text{SM}}=831.7\unit{pb}$ is the SM \ttbar production cross section at next-to-next-to-leading order (NNLO) as calculated with the Top++2.0 program~\cite{Czakon_2014} and $\mathcal{B}(\PQt \to \Pe\PGm\PQq)$ is shown in Eq.~\ref{eq:BR}.

\begin{table*}[tbh]
\centering
\topcaption{Theoretical cross sections for top quark production and decay for each CLFV coupling. The cross sections are calculated with $C_{a}/\Lambda^2=1\TeV^{-2}$, $m_{\PQt} = 172.5\GeV$, $\Gamma_{\PQt}^{\text{SM}} = 1.33\GeV$. The cross section for the top quark decay process is the same for $\Pe\PGm\PQt\PQu$ and $\Pe\PGm\PQt\PQc$ couplings, therefore, only one cross section is quoted for each Lorentz structure. The first uncertainty represents the effect of QCD renormalization and factorization scales. The second uncertainty is the PDF uncertainty.}
\begin{scotch}{lccc}
Signal mode   &  Tensor  & Vector              & Scalar  \\ 
\hline
\rule[-0.2\ruleht]{0pt}{1.15\ruleht}Production ($\Pe\PGm\PQt\PQu$) &  $2140^{+370}_{-290}\pm30$\unit{fb} & $460^{+81}_{-64}\pm6$\unit{fb} & $97^{+18}_{-14}\pm1$\unit{fb}   \\[1mm]
Production ($\Pe\PGm\PQt\PQc$) & $164^{+22}_{-18}\pm27$\unit{fb} & $33^{+5}_{-4}\pm6$\unit{fb}   & $6.3^{+0.9}_{-0.8}\pm1.4$\unit{fb}  \\[1mm]
Decay ($\Pe\PGm\PQt\PQq$, where $\PQq$ = $\PQu/\PQc$) & $187^{+5}_{-6}\pm8$\unit{fb} & $32.0^{+0.8}_{-1.1}\pm1.3$\unit{fb}  & $4.0^{+0.1}_{-0.1}\pm0.2$\unit{fb} \\
\end{scotch}
\label{tab:XS}
\end{table*}

The simulated background processes are divided into three categories: (i) $\PW\PZ$ production; (ii) multiboson production other than $\PW\PZ$ (VV(V), where ${\PV}=\PW, \PZ$), which includes $\PZ\PZ$ and triboson production; and (iii) the associated production of $\PQt(\PAQt)$ with a boson and/or a quark ($\PQt(\PAQt)+{\PX(\PX)}$, where $\PX$ is a boson or quark), which includes $\ttbar \PW$, $\ttbar \PZ$, $\ttbar \PH$, $\PQt\PZ\PQq$, and smaller backgrounds containing one or two top quarks plus a boson or quark.  
Additionally, simulated samples of $\ttbar$ and Drell--Yan (DY) events are used in the data-driven nonprompt-lepton background estimate as well as in the BDT training. 
The \ttbar, $\ttbar \PH$, and $\PZ\PZ$ processes are simulated at next-to-leading order (NLO) precision using the \POWHEG 2.0 event generator~\cite{powheg1, powheg2, powheg3, powheg4}. 
Simulated $\PW\PZ$, $\ttbar \PW$, $\ttbar \PZ$, $\PQt\PZ\PQq$, DY+jets, and triboson events are generated at NLO using \MGvATNLO 2.4.2~\cite{Alwall_2014}. The parton showering and hadronization are modeled using {\PYTHIA}8~\cite{Sjostrand:2014zea}. 

For simulated signal and background events corresponding to the 2016 data-taking period, the NLO PDF set NNPDF3.0~\cite{NNPDF:2014otw} and the {\PYTHIA}8 tune CUETP8M1~\cite{Skands:2014pea} are used. For simulated events matching the conditions of the 2017--2018 data, the NNLO PDF set NNPDF3.1~\cite{NNPDF:2017mvq} and the {\PYTHIA}8 tune CP5~\cite{CMS:2019csb} are used. 
 
The presence of additional $\Pp\Pp$ collisions within the same or adjacent bunch crossings (pileup) is accounted for by overlaying each simulated event with a number of inelastic collisions, simulated using {\PYTHIA}8. Simulated events are weighted such that the distribution of pileup events in the simulated samples matches that observed in the data. All simulated events include a detailed simulation of the CMS detector response based on \GEANTfour~\cite{geant4}, and are processed using the same CMS event reconstruction software as used for the data.

\section{Object reconstruction}
\label{sec:objects}

The particle-flow (PF) algorithm~\cite{CMS:2017yfk} aims to reconstruct and identify each individual particle in an event, with an optimized combination of information from the various elements of the CMS detector, including charged particle tracks from the tracking detector, energy deposits in the HCAL and ECAL, and reconstructed tracks from the muon chambers. Particles in each event are reconstructed and identified as either electrons, muons, photons, charged hadrons, or neutral hadrons. 
The primary interaction vertex is taken to be the vertex corresponding to the hardest scattering in the event, evaluated using tracking information alone, as described in Section 9.4.1 of Ref.~\cite{CMS-TDR-15-02}.

Photons are identified as ECAL energy clusters not linked to the extrapolation of any charged particle trajectory. Electrons are identified as a primary charged particle track and potentially multiple ECAL energy clusters corresponding to the extrapolation of this track to the ECAL and to possible bremsstrahlung photons emitted along the way. Muons are identified as tracks in the tracker consistent with either a track or several hits in the muon system, and associated with calorimeter deposits compatible with the muon hypothesis. Charged hadrons are identified as charged particle tracks neither identified as electrons nor as muons. Finally, neutral hadrons are identified as HCAL energy clusters not linked to any charged hadron candidates, or as a combined ECAL and HCAL energy excess with respect to the expected charged hadron energy deposit. 

The energy of electrons is determined from a combination of the electron momentum measurement in the tracker at the primary interaction vertex, the energy of the corresponding ECAL cluster, and the energy sum of all bremsstrahlung photons spatially compatible with originating from the electron track. The energy of muons is obtained from the curvature of the corresponding track. The energy of charged hadrons is determined from a combination of their momentum measured in the tracker and the matching ECAL and HCAL energy deposits, corrected for the response function of the calorimeters to hadronic showers. Finally, the energy of neutral hadrons is obtained from the corresponding corrected ECAL and HCAL energy deposits. 

The PF candidates are clustered into jets using the anti-$\kt$ clustering algorithm~\cite{Cacciari:2008gp} with a distance parameter of 0.4, implemented in the \FASTJET package~\cite{Cacciari:2011ma,Cacciari:2005hq}. The jet momentum is determined as the vectorial sum of all particle momenta in the jet, and is found from simulation to be, on average, within 5--10\% of the true momentum over the entire transverse momentum (\pt) spectrum and detector acceptance. 
Additional pileup interactions can contribute additional tracks and calorimetric energy depositions, increasing the apparent jet momentum. To mitigate this effect, tracks identified as originating from pileup vertices are discarded and a correction is applied for the remaining contributions. 
Jet energy corrections are derived from simulation studies so that the average measured energy of reconstructed jets becomes identical to that of particle-level jets. In situ measurements of the momentum balance in dijet, photon+jet, $\PZ$+jet, and multijet events are used to determine any residual differences between the jet energy scale in data and in simulation, and appropriate corrections are made~\cite{CMS:2016lmd}. Additional selection criteria are applied to each jet to remove jets potentially dominated by instrumental effects or reconstruction failures~\cite{CMS:JME-16-003}. 

The missing transverse momentum vector \ptvecmiss is computed as the negative vector \pt sum of all the PF candidates in an event, and its magnitude is denoted as \ptmiss~\cite{CMS:2019ctu}. The \ptvecmiss is modified to account for corrections to the energy scale of the reconstructed jets in the event.

\section{Event selection}
\label{sec:selection}

Electrons and muons used in the analysis are identified using a set of kinematic and quality selection criteria, as well as a multivariate analysis (MVA) discriminant derived from a BDT, described in Ref.~\cite{CMS:2021ugv}. All leptons are required to fulfill $\pt > 20\GeV$ and $\abs{\eta}<2.4$. Electrons that are reconstructed within the transition region between the barrel and the endcap sections of the ECAL, $1.44 < \abs{\eta} < 1.57$, are excluded from the analysis as the reconstruction of an electron object in this region is suboptimal. The BDT uses multiple discriminating observables, such as lepton kinematic and isolation variables. It is trained on simulated $\PQt\PAQt\PW$, $\PQt\PAQt\PZ$, $\PQt\PZ\PQq$, and $\ttbar$ events, and is optimized to reduce the contamination from nonprompt leptons. Both electrons and muons are required to pass the tight MVA working point, corresponding to 90\% (1\%) signal (background) efficiency.

A lepton isolation variable is defined as the \pt sum of all reconstructed PF particles within $\Delta R = \sqrt{\smash[b]{(\Delta \eta)^2 + (\Delta \phi)^2}} < R_{\max}$ (where $\Delta \phi$ is the azimuthal angle difference in radians) around the direction of the lepton, excluding the lepton itself. The cone size $R_{\max}$ depends on the lepton \pt as $R_{\max} = \max(0.05,\min(0.2,\frac{10\GeV}{\pt}))$. The value of the isolation variable divided by the lepton \pt, referred to as relative isolation, is required to be less than 0.12. Leptons selected by these requirements are referred to as \emph{tight} leptons.

A selection of \emph{loose} leptons is also used in the estimation of the nonprompt background. The loose lepton criteria require leptons to pass a much looser MVA working point, corresponding to 99\% (10\%) and 98\% (5\%) signal (background) efficiency for electrons and muons, respectively, and to have a relative isolation less than 0.4. Unless explicitly stated, all leptons must pass the tight criteria.

Jets are required to have $\pt > 30\GeV$, $\abs{\eta}<2.4$, and be separated from the selected leptons by $\Delta R>0.4$. An MVA-based discriminant (\textsc{DeepJet}) is used to identify jets that originate from the hadronization of bottom quarks~\cite{BTV-16-002, Bols:2020bkb, CMS-DP-2018-058}. The working point is chosen such that the \PQb jet selection efficiency is 75\% while the rate of misidentified light-quark or gluon jets as \PQb jets is 1\%~\cite{CMS-DP-2018-058}.

Events selected for the analysis are required to contain exactly three leptons with a net charge of $\pm1$. The leading (highest \pt) lepton must have $\pt > 38\GeV$ and be matched with the corresponding object used by the trigger within $\Delta R<0.2$. 
Events are further divided into three categories (\eee, \mmm, and \eml), and within each category different regions are defined to further study and validate signal and background events---these are described below.

\subsection{Event categories}

Events with three leptons of the same flavor are categorized into the \eee or \mmm category and are used to study the background composition. 
Remaining events are categorized into the \eml category, which is where the signal is expected; for this category, we require at least one pair of opposite-sign, different-flavor (OSDF) leptons. The OSDF lepton pair is assumed to be the product of the CLFV interaction, and is referred to as the LFV $\Pe\PGm$ pair. Leptons that form the LFV $\Pe\PGm$ pair are referred to as the LFV electron/muon. The third lepton (referred to as the standalone lepton), together with a neutrino, are assumed to be the products of the leptonically decaying $\PW$ boson, which arises from the SM top quark decay -- the lepton assignments are described below.   

A kinematic reconstruction of the neutrino momentum is performed, which proceeds as follows. The $x$ and $y$ components of the neutrino momentum are taken from \ptvecmiss, while the $z$ component ($p_z$) is calculated by imposing the constraint that the invariant mass of the standalone lepton and the neutrino must equal the $\PW$ boson mass. If there is no real solution for $p_z$, we take the real part of the complex solution and if there is more than one real solution, we select the one closest to the $p_z$ of the standalone lepton.

The jet with the highest $\PQb$ tagging score is assumed to originate from a bottom quark decay and is combined with the standalone lepton and the neutrino to construct the SM top quark candidate. 
In events where there is more than one candidate for the standalone lepton, the lepton that gives a top quark candidate mass closest to the nominal top quark mass ($m_{\PQt}=172.5\GeV$) is chosen.
Once the standalone lepton has been determined, the OSDF pair is combined with each selected jet in the event (other than the one with the highest $\PQb$ tagging score) to reconstruct potential LFV top quark candidates. From the LFV top quark candidates, the one with mass closest to $m_{\PQt}$ is chosen. For events with only one jet (used to reconstruct the SM top quark), no LFV top quark candidate is formed. 

\subsection{Signal region}

The signal region (SR) targeting LFV processes is defined from events selected in the \eml category. In addition to the OSDF lepton pair, we require the events to have $\ptmiss> 20\GeV$. Events with an opposite-sign, same-flavor (OSSF) lepton pair with an invariant mass between 50 and 106\GeV are removed (OffZ requirement) to suppress contributions from $\PZ$ boson production in association with jets. To further suppress background events, we require the presence of at least one jet of which at most one is $\PQb$ tagged. 
The selection criteria used to define the SR are summarized in Table~\ref{tab:region}. 

Since the invariant mass of the opposite-sign $\Pe\PGm$ pair in the top quark production signal is not bounded by $m_{\PQt}$, the SR is further subdivided to create regions enriched in either top quark decay (Fig.~\ref{fig:Feynman_Diagram} left) or production (Fig.~\ref{fig:Feynman_Diagram} middle and right) signal events:
\begin{itemize}
\item $\text{SR}+m({\Pe\PGm})<150\GeV$: top quark decay enriched,
\item $\text{SR}+m({\Pe\PGm})>150\GeV$: top quark production enriched.
\end{itemize}

\subsection{Signal-depleted regions}

We define a nonprompt background validation region (VR) within each event category (\eee/\mmm VR and \eml VR). For the \eml VR, at least one OSSF lepton pair with an invariant mass between 50 and 106\GeV is required (OnZ) to ensure orthogonality with the SR. 
We also define a $\PW\PZ$ VR in each category---for these VRs at least one OnZ OSSF lepton pair with $\ptmiss> 20\GeV$ is required. The selection criteria associated with each event region are listed in Table~\ref{tab:region}. The VRs are further discussed in Section~\ref{sec:background}. The nonprompt and WZ VRs are only used to validate the modeling of the background and are not included in the statistical analysis described in Section~\ref{sec:results}.

A modified version of the $\PW\PZ$ VR, the jet modeling VR, is used to study the modeling of the jet multiplicity for the diboson production processes. The jet modeling VR has the same OnZ requirement, no jet multiplicity requirement, a $\ptmiss>85\GeV$ requirement, and a requirement of no b-tagged jets with $\pt>20\GeV$. Unlike for the $\PW\PZ$ VR, events with different lepton flavor compositions are combined. 

\begin{table*}[tbh]
\centering
\topcaption{Summary of the selection criteria used to define different event regions. The jet modeling VR contains events from all three categories (\eee, \eml, \mmm).}
\begin{scotch}{ccccccc}
Category         &Region & OnZ & OffZ & $\ptmiss$ & $\# \, \text{jets} $ & $\# \, \PQb \, \text{jets}$\\ 
\hline
                       3$\Pell$      & jet modeling VR & \checkmark      & \NA       & $>$85\GeV       & \NA     & $=$0   \\ 
\hline 
\multirow{2}{*}{\eee/\mmm}     & nonprompt VR & \NA       & \NA       & \NA       & \NA     & \NA   \\  
            & $\PW\PZ$ VR & \checkmark   & \NA       &  $>$20\GeV   & $\geq$1 & $\leq$1\\ 
\hline
\multirow{3}{*}{\eml}   & SR  & \NA       & \checkmark   &  $>$20\GeV   & $\geq$1 &  $\leq$1 \\
            & nonprompt VR  & \checkmark   & \NA       & \NA       & \NA     & \NA     \\
            & $\PW\PZ$ VR & \checkmark   & \NA       &  $>$20\GeV   & $\geq$1 &  $\leq$1 \\
\end{scotch}
\label{tab:region}
\end{table*}

\section{Nonprompt background estimation}
\label{sec:background}

Different background processes contribute to the SR. We distinguish between events with three prompt leptons and events containing nonprompt leptons. 
In this analysis, the term \emph{prompt} leptons include leptons that originate from the CLFV vertex, the DY process, or an electroweak boson decay, including leptons from $\Pgt$ decays if the $\Pgt$ originates from the latter two processes. \emph{Nonprompt} leptons include leptons from hadron decays and photon conversions, as well as particles misidentified as leptons. Nonprompt leptons are suppressed through isolation requirements and an MVA-based identification specifically trained to reject them.

The nonprompt background consists of events with at least one nonprompt lepton passing the analysis selection, and is dominated by DY and $\ttbar$ production. An accurate estimation of the nonprompt-background yield is difficult to achieve through MC simulation. Therefore, a technique based on control samples in data, known as the \emph{matrix method} (described, \eg, in Ref.~\cite{matrixmethod}) is used to estimate this background contribution. 

\subsection{The matrix method}

The matrix method estimates the number of events with nonprompt leptons passing the tight lepton selection. 
We consider a maximum of three nonprompt leptons in the event. 
The matrix method requires knowledge of the probability for prompt and nonprompt leptons that pass a loose lepton selection to also pass a tight lepton selection. The loose selection is such that any leptons that pass it will fall into one of two exclusive categories: tight or not tight. 
The tight and loose selections are detailed in Section~\ref{sec:selection}. 

The prompt (real) and nonprompt (fake) rates ($r$ and $f$) measure the probability for prompt and nonprompt leptons, respectively, to pass the tight selection.  
The measurements of $r$ and $f$ are performed in dedicated dilepton VRs, where a high purity of prompt and nonprompt leptons, respectively, is expected. The leading lepton is selected with the \emph{tight} criteria while the sub-leading lepton is selected with the \emph{loose} criteria. Events in these dilepton VRs are further required to contain at least one jet of which no more than one is associated with a bottom quark. 

The prompt rate $r$ is measured using $\ttbar$ simulation in an opposite-sign dilepton region, where only $\Pe\PGm$ events are included. The nonprompt rate $f$ is measured using data in a same-sign dilepton region, where all lepton-flavor composites ($\Pe\Pe$, $\Pe\PGm$, and $\PGm\PGm$) are included. Both $r$ and $f$ are measured separately for electrons and muons, and are parameterized as functions of \pt, $\abs{\eta}$, and jet multiplicity. 
It is assumed that $r$ and $f$ are local properties of the lepton identification, which are largely independent of the event-level quantities. Nevertheless, the variation of $f$ for nonprompt leptons produced in different processes can be quite significant. The dependency of $f$ on the physics processes can manifest when different lepton flavors are required: the same-flavor dilepton VRs ($\Pe\Pe$, $\PGm\PGm$) are mostly populated by DY events while \ttbar is the primary contributor to the different-flavor ($\Pe\PGm$) dilepton VR. The dependency of $f$ measured with different lepton-flavor composites, which cannot be fully captured by our parametrization, is covered by introducing a dedicated uncertainty, corresponding to a shift of $f$ by up to 20\%.

The nonprompt background estimate in a given event region (\eg, the SR) is determined with the matrix method using a number of application regions (ARs). The event-level selections for these ARs are the same as for the event region in question, except that all three leptons are selected by loose lepton selection criteria. Since the loose leptons selected in the ARs can either pass or fail the tight lepton requirement, the ARs are categorized based on the composition of tight/not tight leptons. 
Events in each AR are reweighted according to weights produced by the matrix method using $r$ and $f$ to obtain the nonprompt background estimate in a given region.

\subsection{Validation of the matrix method}

\begin{figure*}[tbhp]
 \centering
 \includegraphics[width=0.38\textwidth]{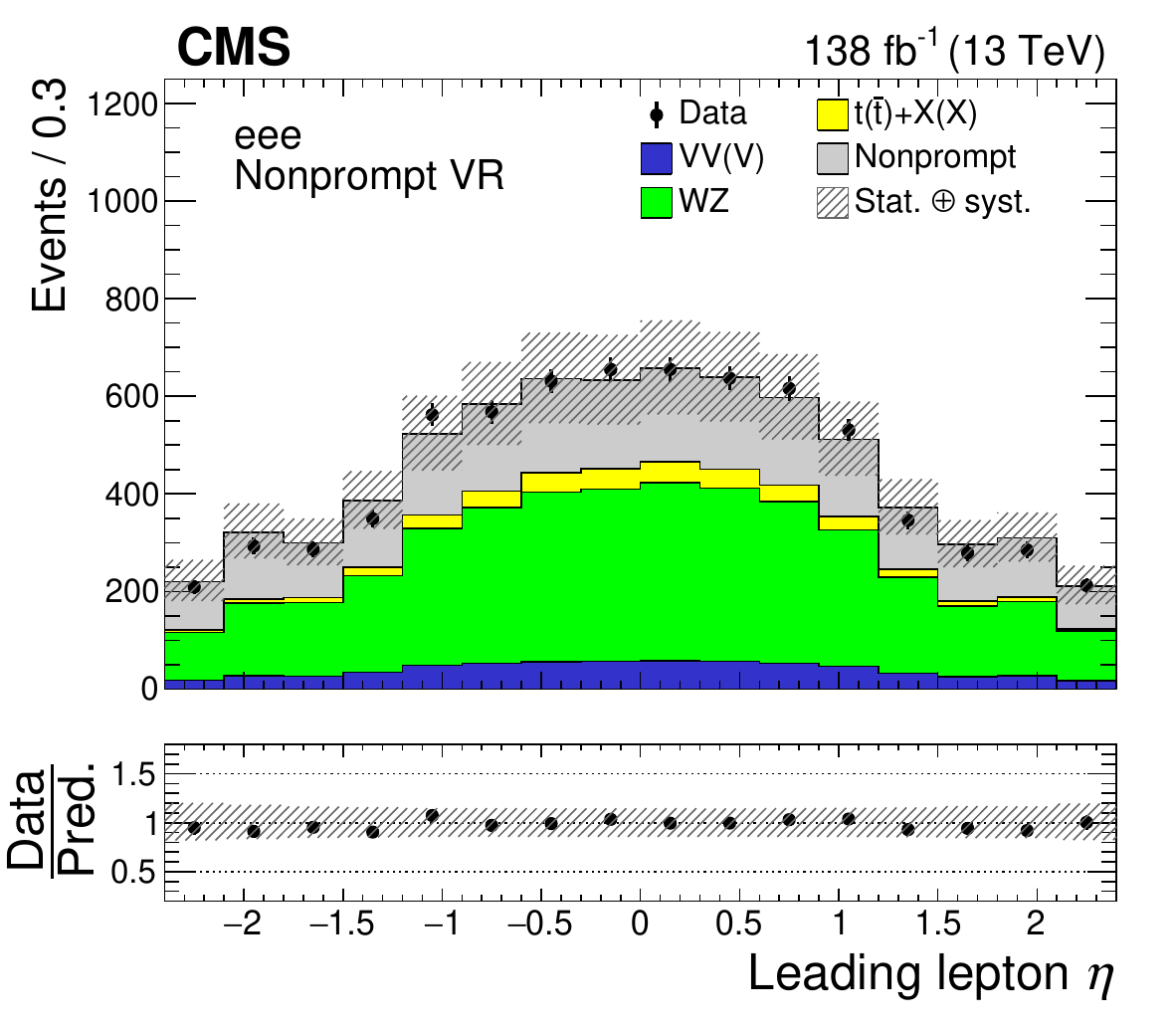}
 \includegraphics[width=0.38\textwidth]{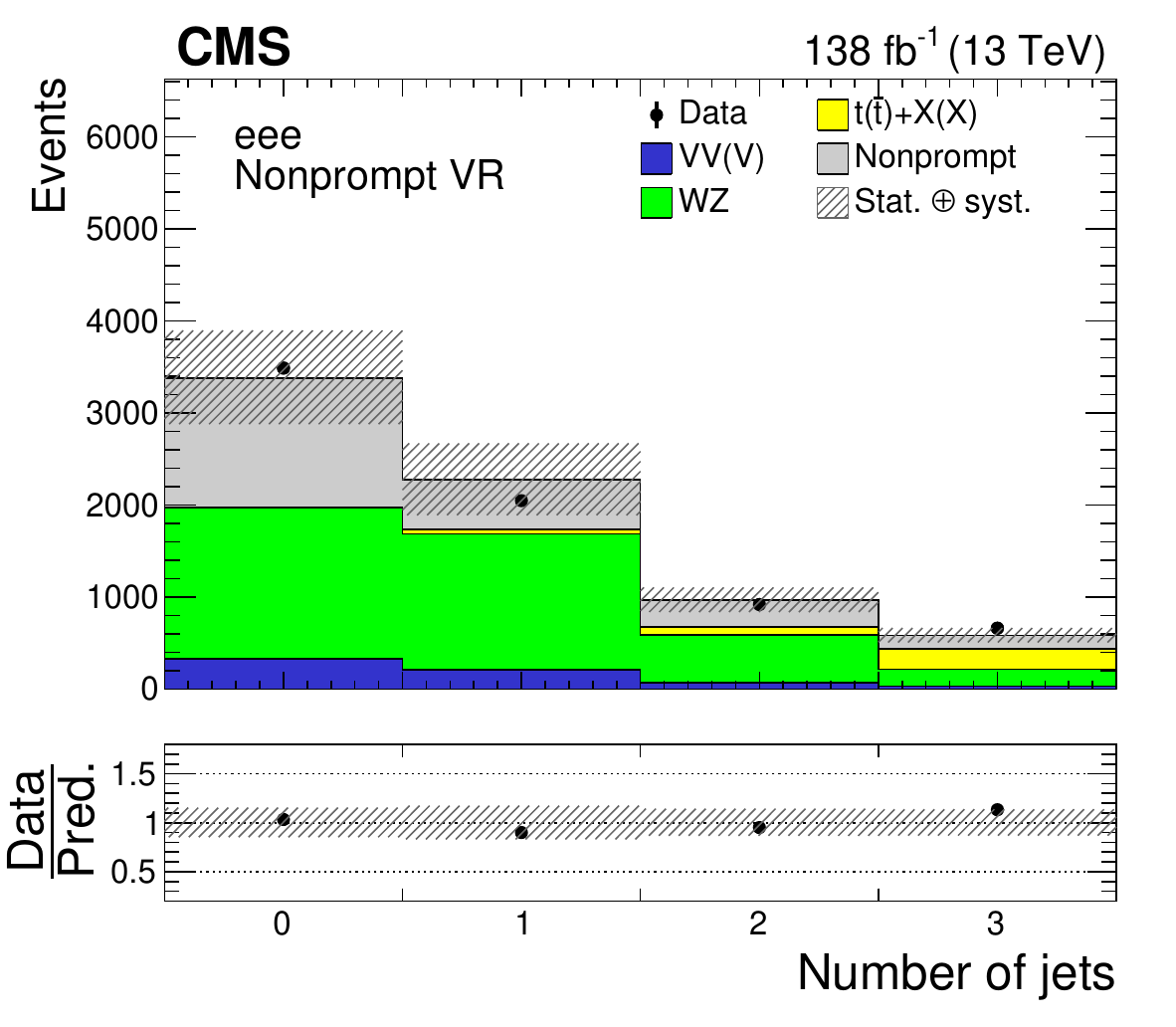}
 \includegraphics[width=0.38\textwidth]{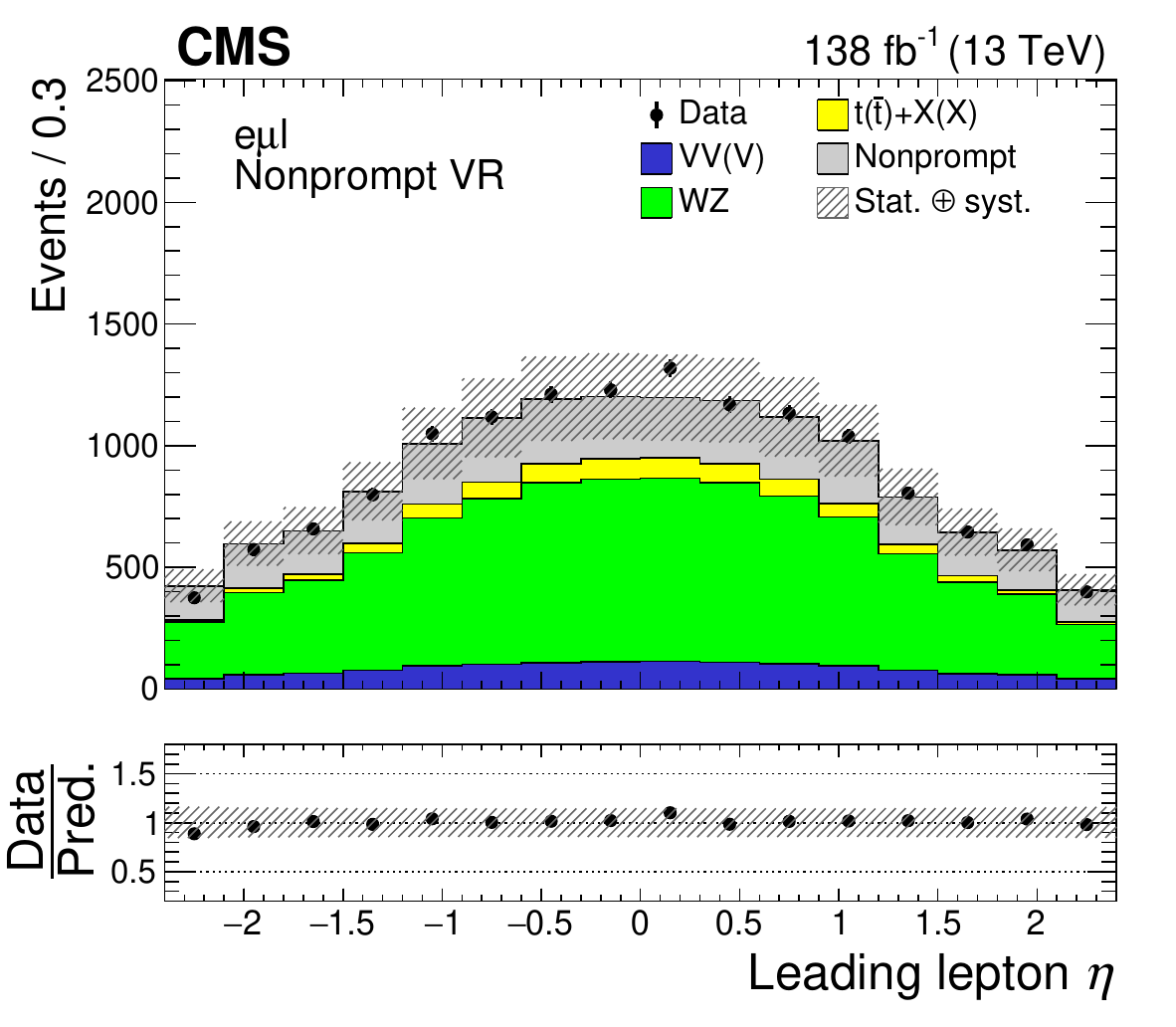}
 \includegraphics[width=0.38\textwidth]{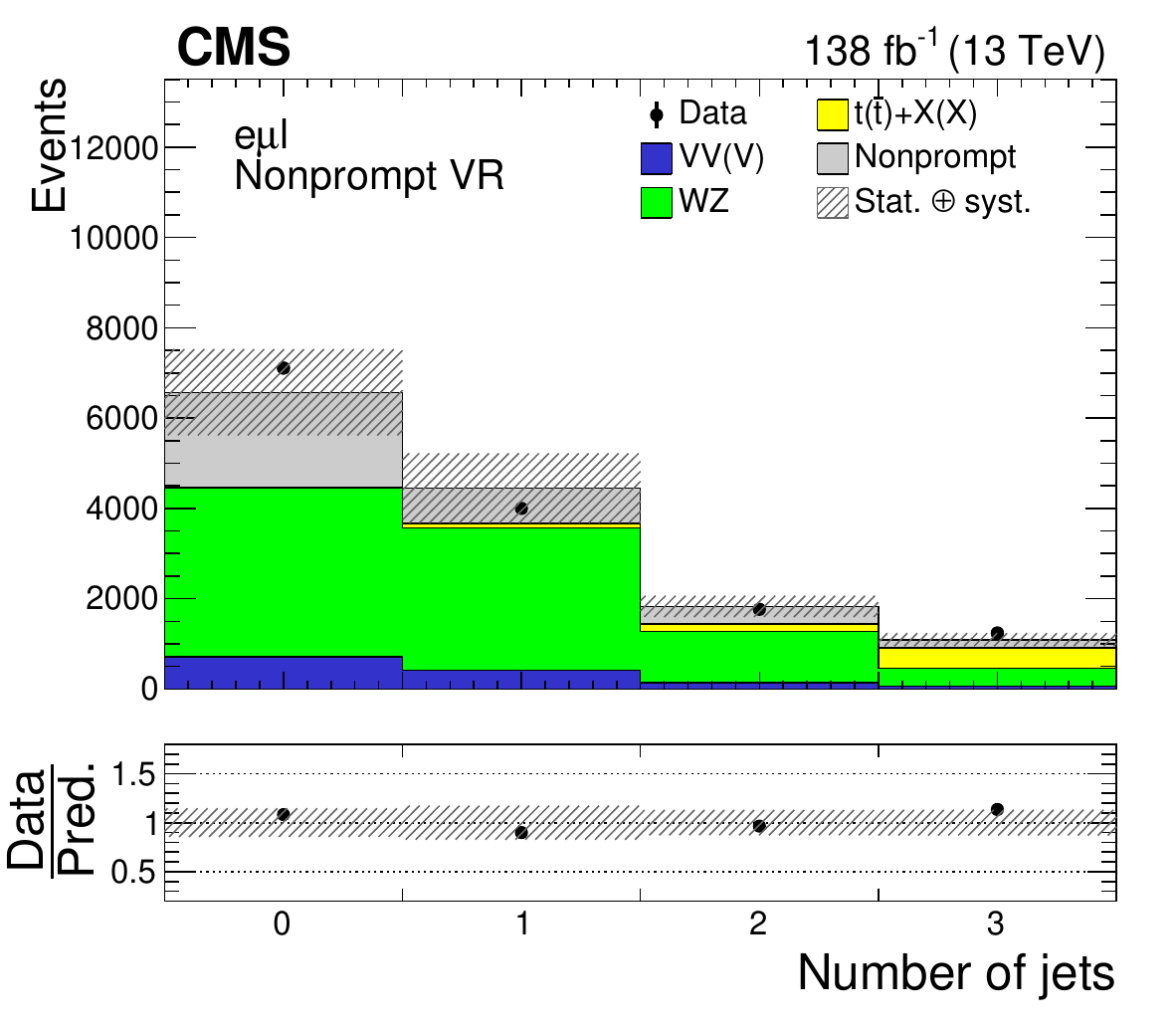} 
 \includegraphics[width=0.38\textwidth]{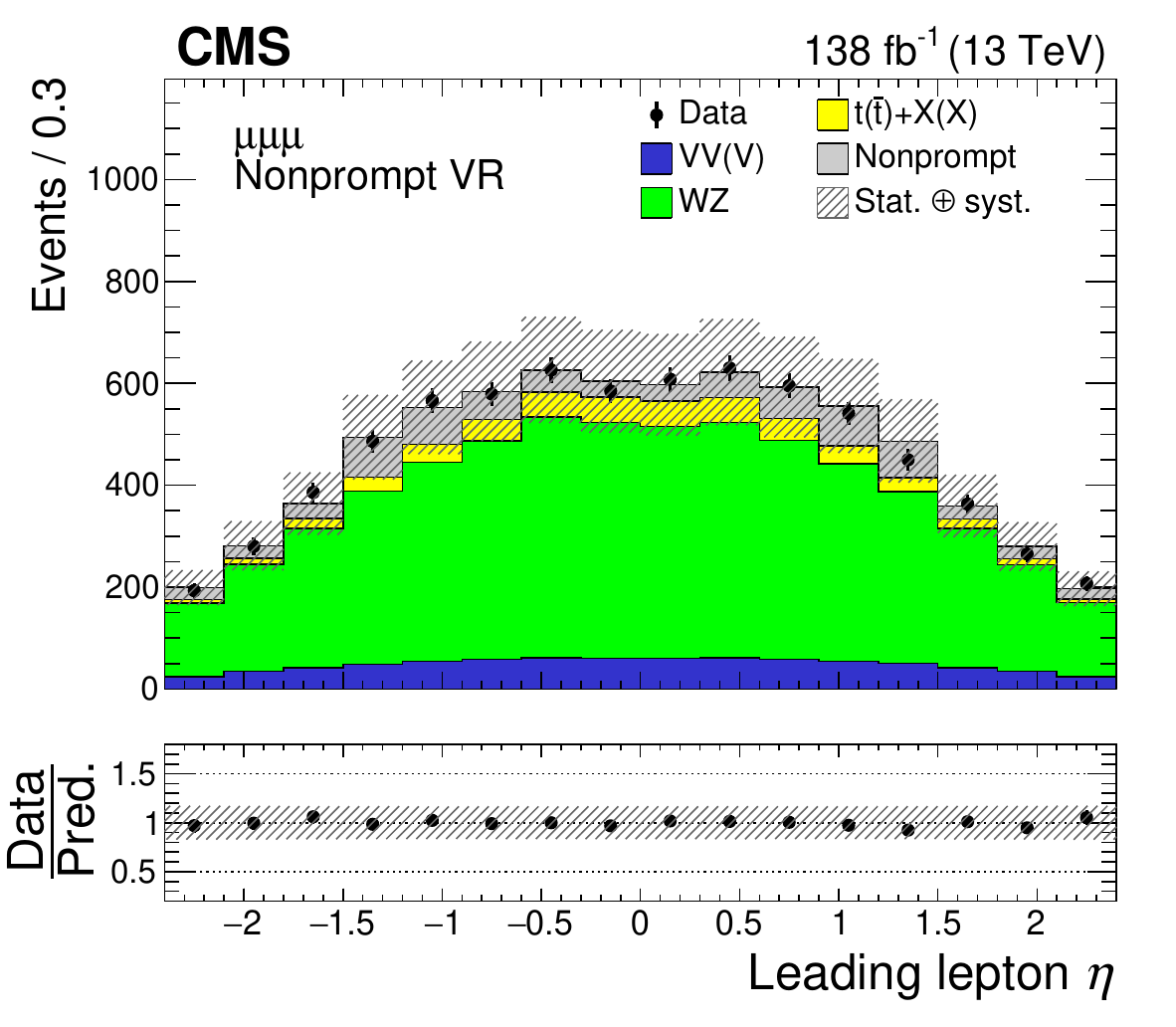}
 \includegraphics[width=0.38\textwidth]{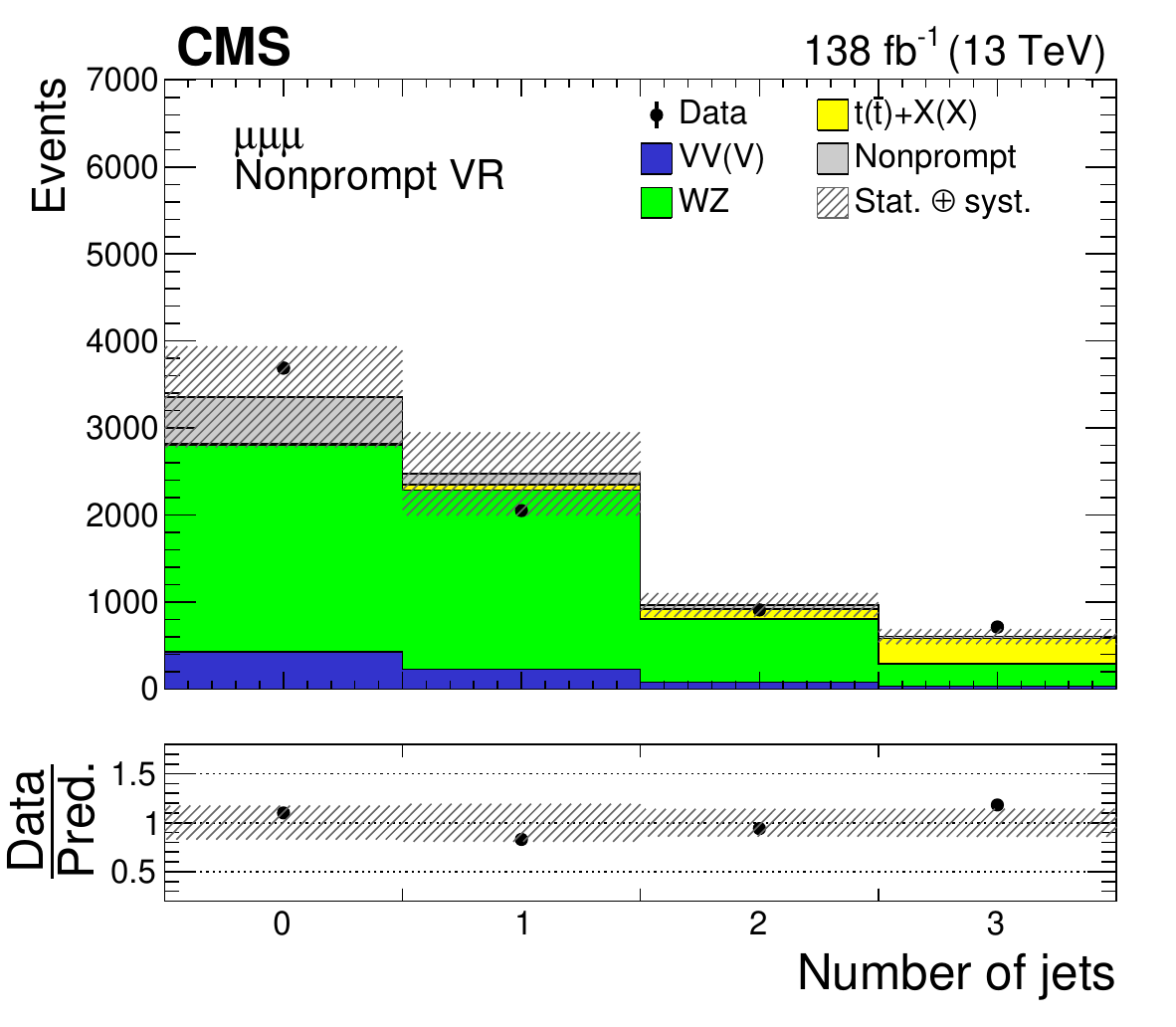} 
 \caption{Distributions of the leading lepton $\eta$ (left) and the jet multiplicity (right) in the nonprompt VRs. Events in the \eee, \eml, and \mmm nonprompt VRs are shown in the upper, middle, and lower row, respectively. The data are shown as filled points and the SM background predictions as histograms. 
The VV(V) background includes $\PZ\PZ$ and triboson production, while the $\PQt(\PAQt)+{\PX(\PX)}$ component includes $\ttbar \PW$, $\ttbar \PZ$, $\ttbar \PH$, $\PQt\PZ\PQq$, and smaller backgrounds containing one or two top quarks plus a boson or quark.  
The nonprompt background is estimated using control samples in data, while other backgrounds are estimated using MC simulation. The hatched bands indicate statistical and systematic uncertainties in the background predictions. The last bin of the right column histograms includes the overflow events.}
 \label{fig:VR_matrix}
\end{figure*}

Three nonprompt VRs that are orthogonal to the SR are used to validate the performance of the matrix method. The selection criteria used to define the VRs are summarized in Section~\ref{sec:selection}. For each VR, the prompt backgrounds are estimated with MC simulation, while the nonprompt background is estimated with the matrix method.
The distribution of the leading lepton $\eta$ and the jet multiplicity are shown in Fig.~\ref{fig:VR_matrix} for the three nonprompt VRs (\eee, \eml, and \mmm VRs). Good agreement is observed between the data and background estimates, serving as a validation of the matrix method. 

\subsection{Signal region background estimation}   

\begin{figure*}[tbhp]
 \centering
 \includegraphics[width=0.38\textwidth]{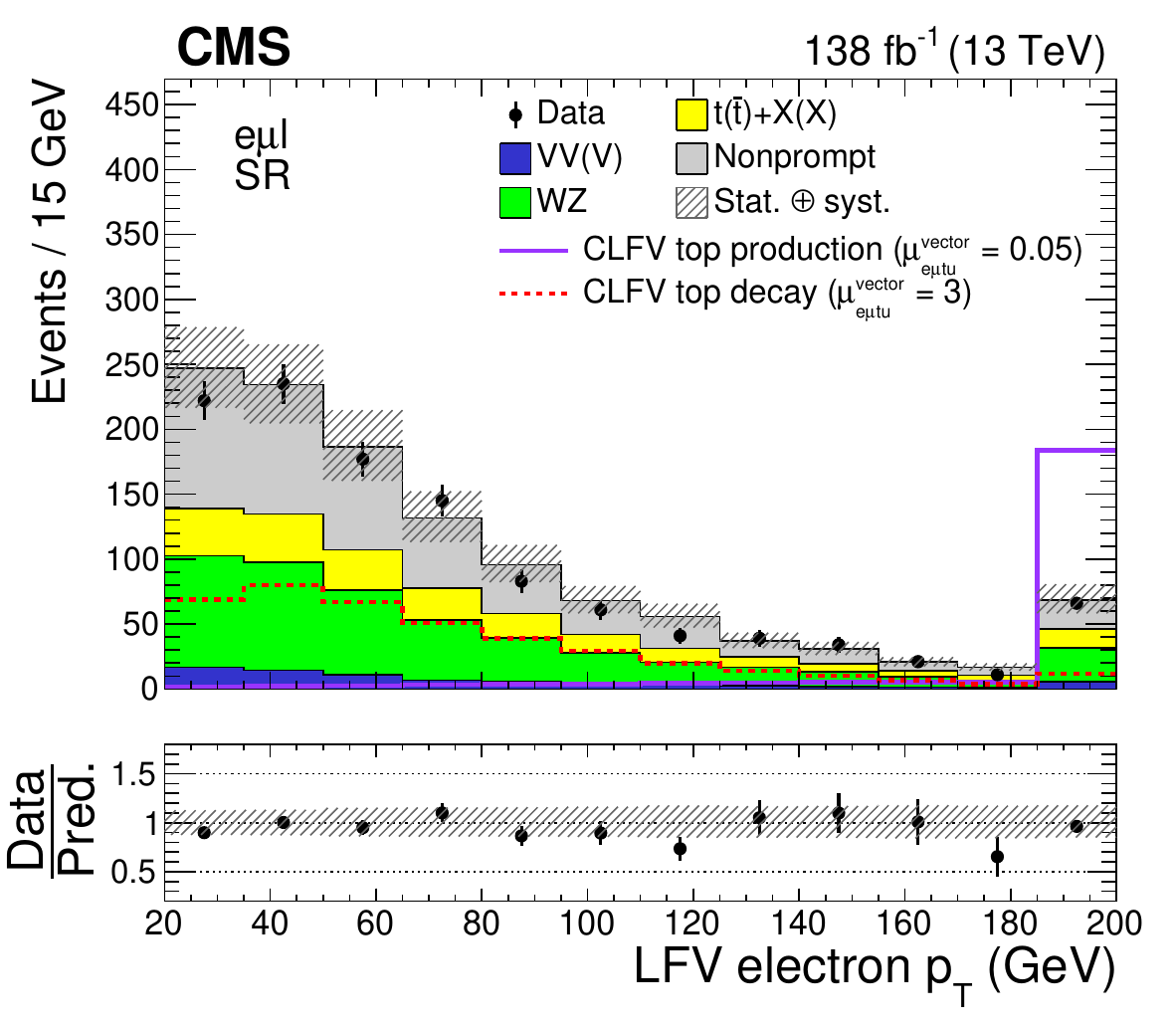}
 \includegraphics[width=0.38\textwidth]{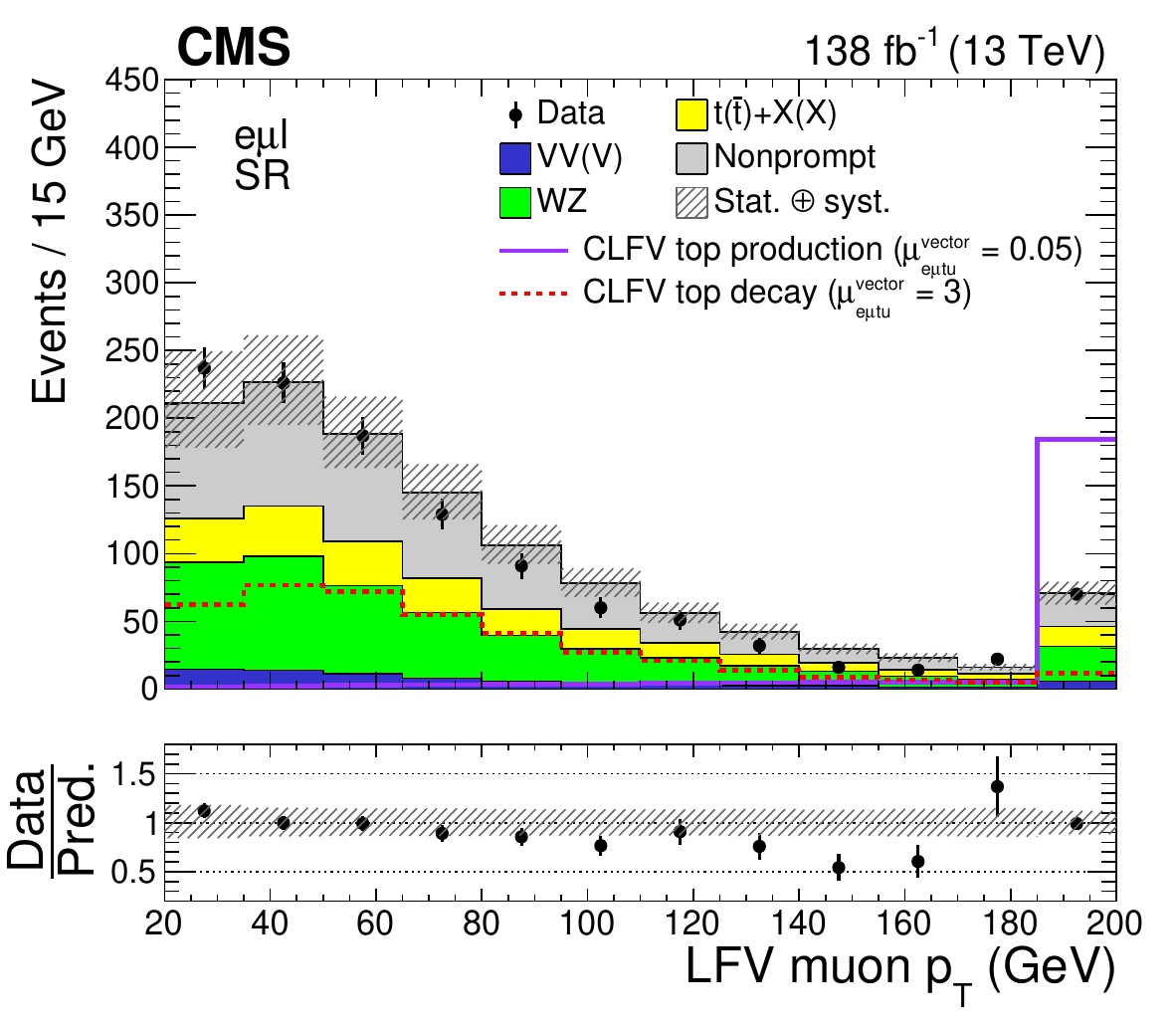}
 \includegraphics[width=0.38\textwidth]{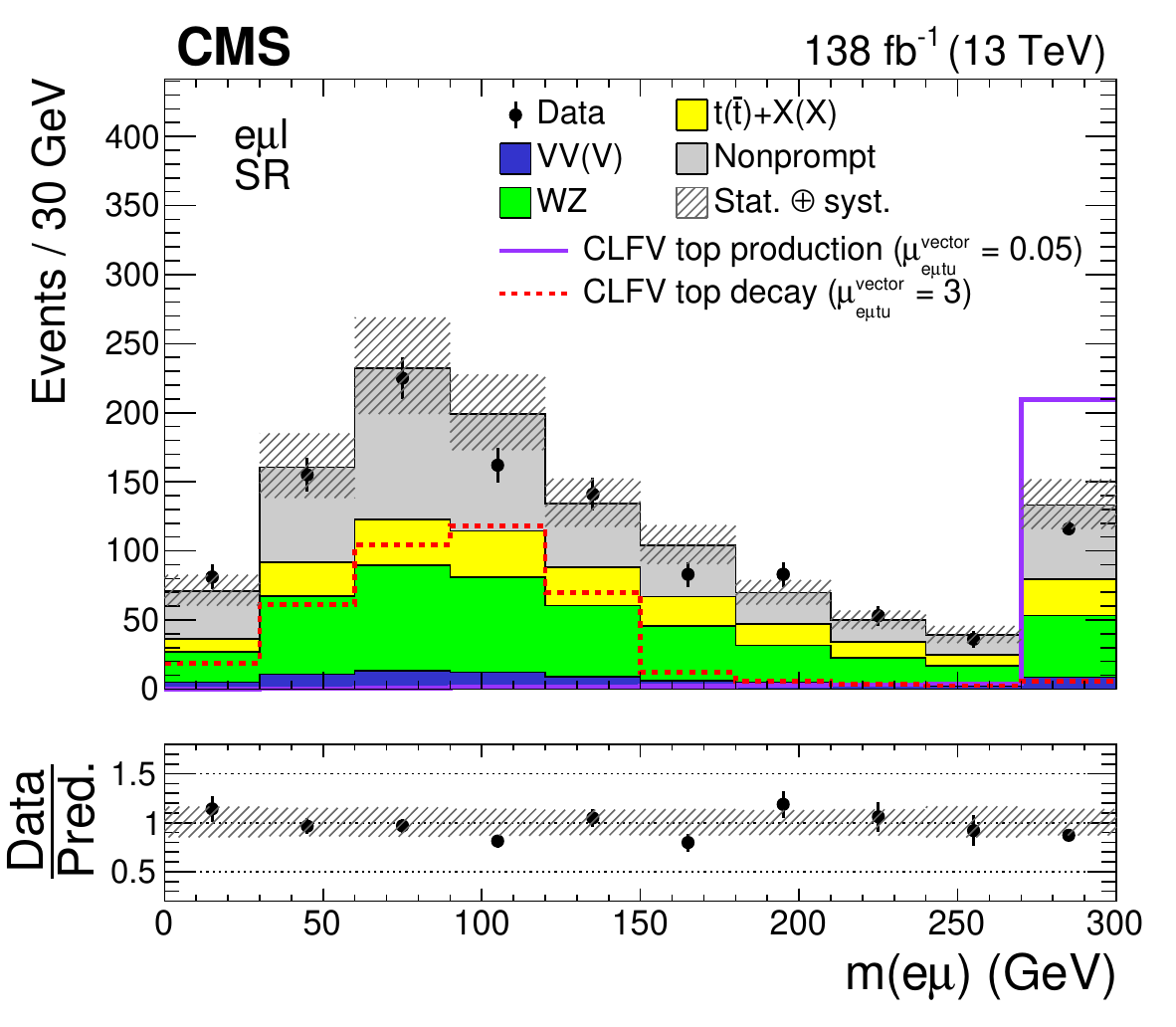}
 \includegraphics[width=0.38\textwidth]{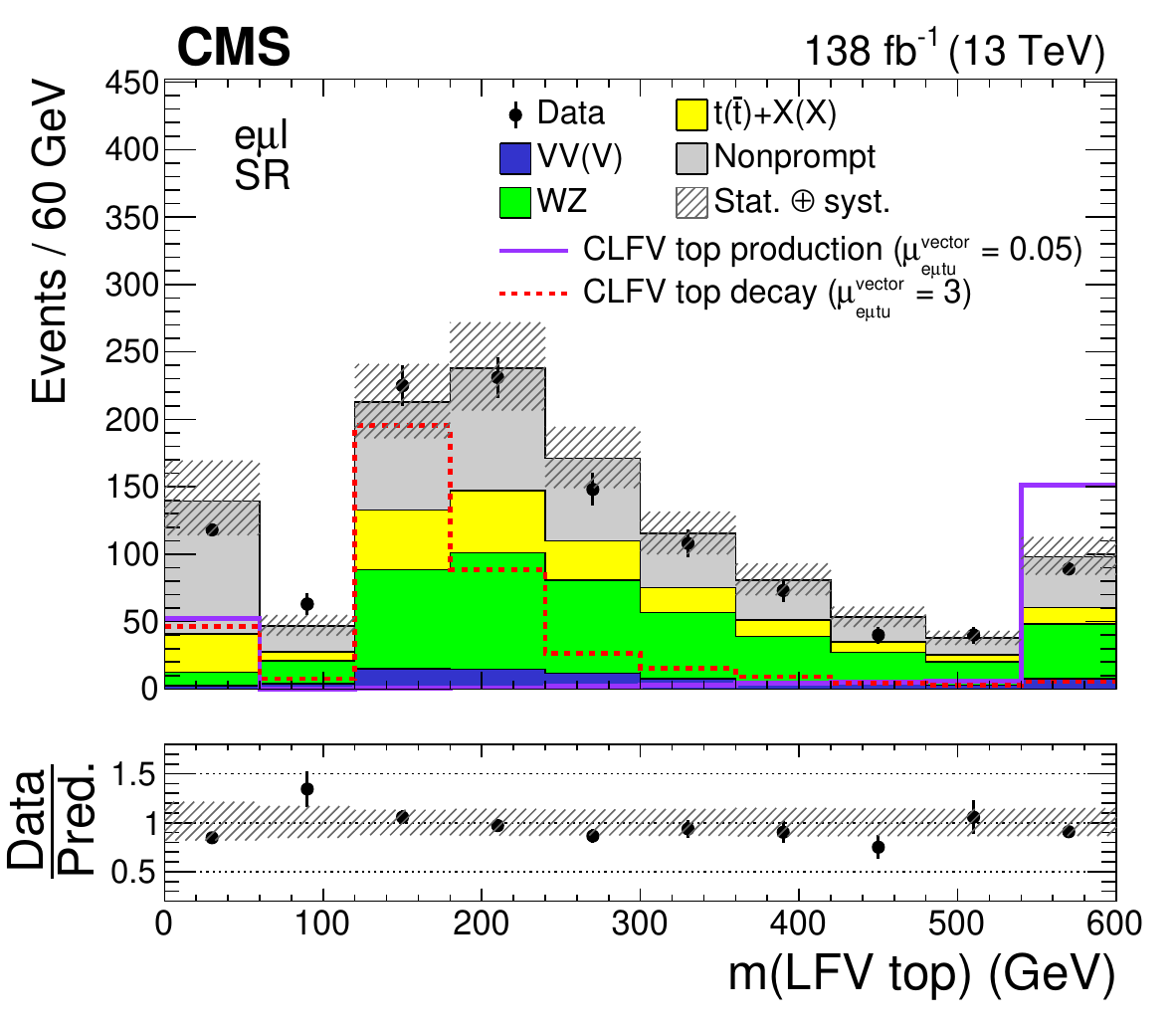} 
 \includegraphics[width=0.38\textwidth]{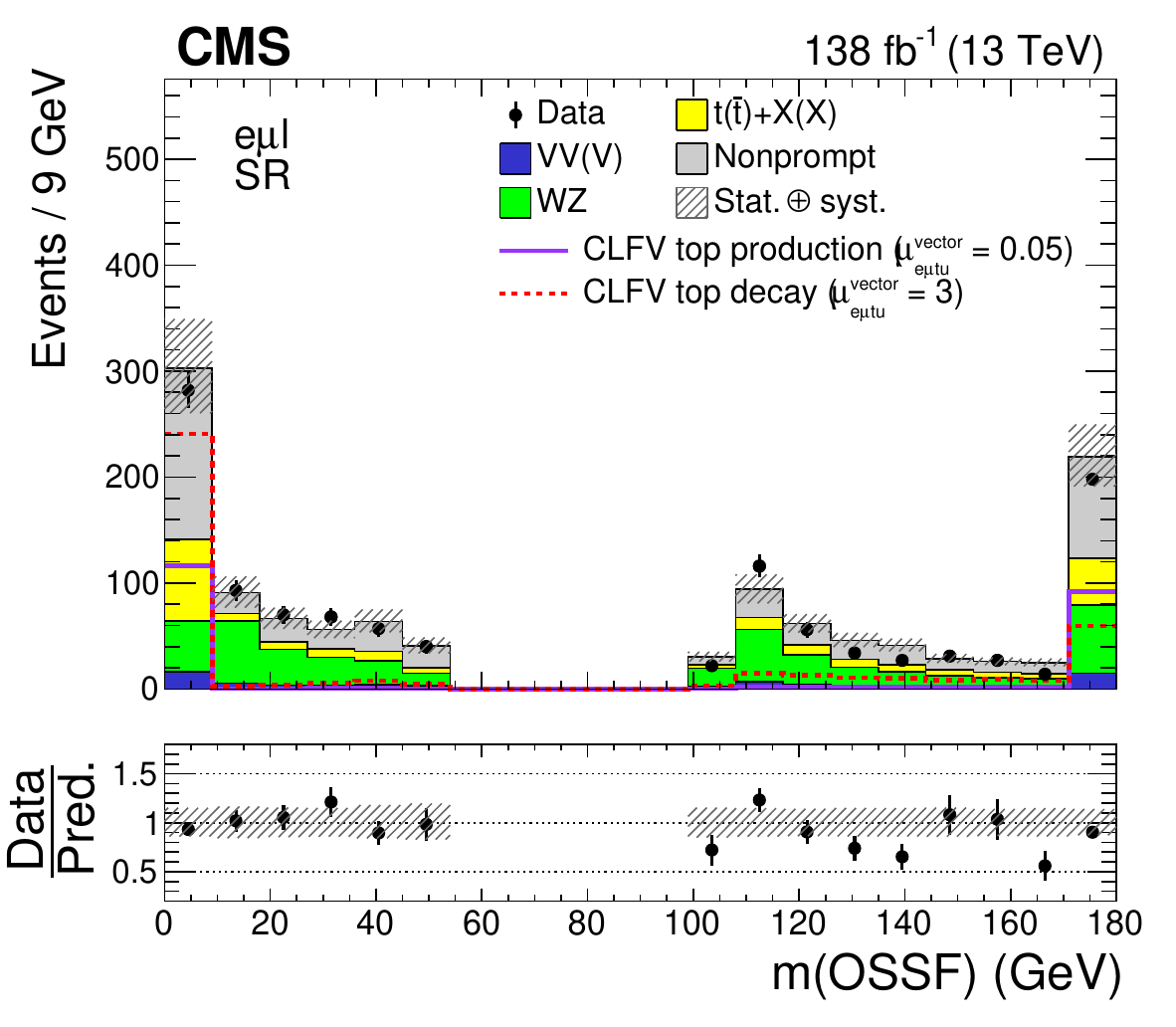}
 \includegraphics[width=0.38\textwidth]{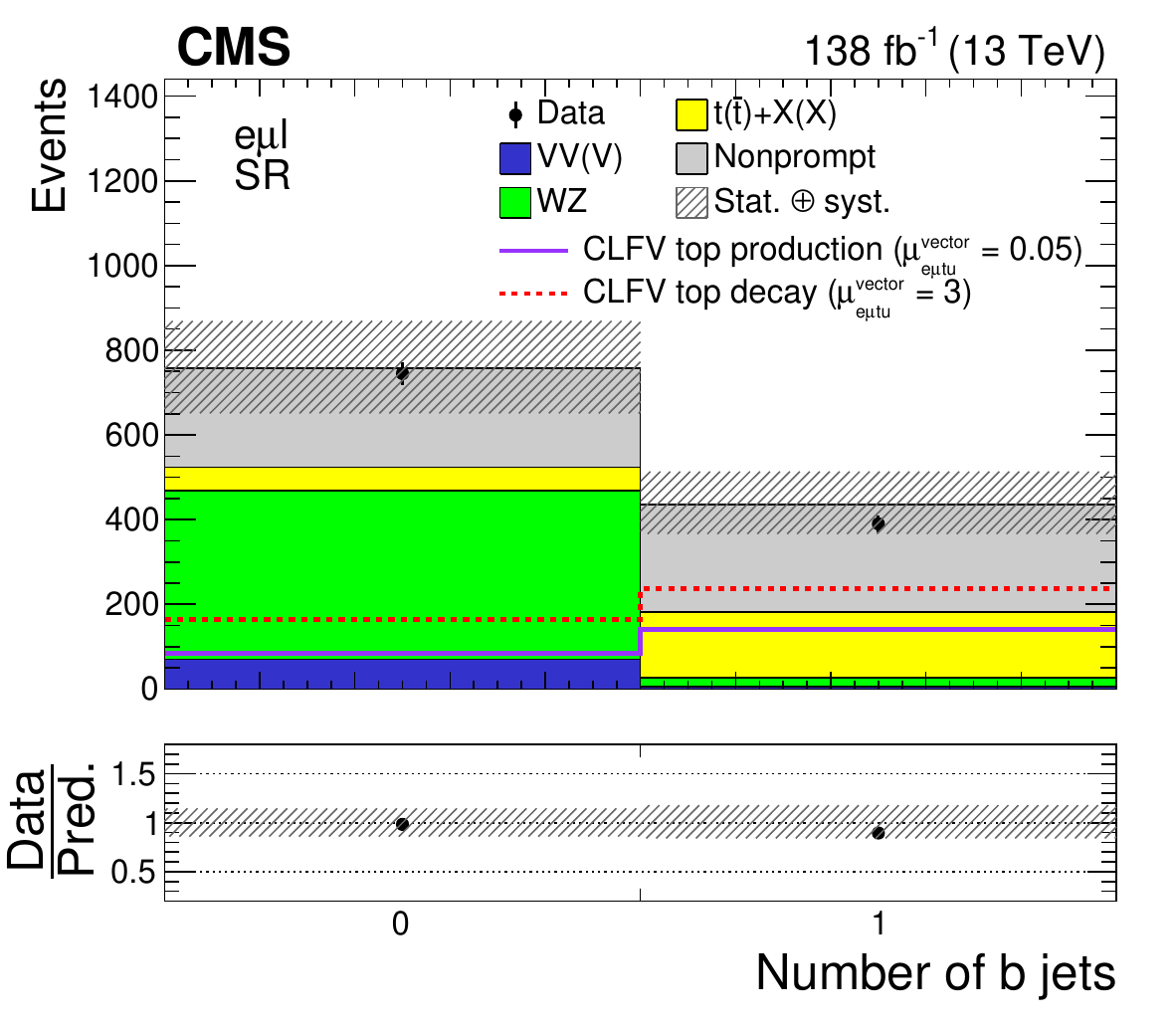}
 \caption{Distributions of kinematic variables in the SR: LFV electron \pt (upper left), LFV muon \pt (upper right), LFV $\Pe\PGm$ mass (middle left), LFV top quark mass (middle right), OSSF lepton pair mass (lower left), and $\PQb$ jet multiplicity (lower right). The CLFV top quark decay and production signals are shown as dotted red and solid purple lines, respectively. The original signal normalization, corresponding to $C_{\Pe\PGm\PQt\PQu}^{\text{vector}}/\Lambda^2=1\TeV^{-2}$, is scaled up (down) by a factor of 3 (20) for the CLFV top quark decay (production) signal for better visualization. The hatched bands indicate statistical and systematic uncertainties in the background predictions. The last bin of all but the lower-right histogram includes the overflow events.}
 \label{fig:SR_DataDriven}
\end{figure*}

The matrix method is used to estimate the nonprompt background in the SR. Distributions of LFV electron \pt, LFV muon \pt, LFV $\Pe\PGm$ mass, LFV top quark mass, OSSF boson mass, and $\PQb$ jet multiplicity are shown in Fig.~\ref{fig:SR_DataDriven}. In these plots, the signal strength $\mu$ describes the scaling of the signal cross section as:
\begin{equation}
\mu(C/\Lambda^2)=\frac{\sigma_{\text{CLFV}}(C/\Lambda^2)}{\sigma_{\text{CLFV}}(1\TeV^{-2})}\propto(C/\Lambda^2)^2.
\label{eq:signal_strength}
\end{equation}
Signal events generated with the vector-like operator that features an up quark in the CLFV coupling are used as an example in these plots to highlight the difference between signal and background in various distributions. Event yields for the various background processes are summarized in Table~\ref{tab:eventcount}. The data are observed to agree well with the background estimate. 

\begin{table}[tbh]
\centering
\topcaption{Expected background contributions and the number of events observed in data collected during 2016--2018. The quoted uncertainties include statistical and systematic sources, which are added in quadrature. The category ``Other" includes smaller background contributions containing one or two top quarks plus a boson or quark. The CLFV signal, generated with $C_{\Pe\PGm\PQt\PQu}^{\text{vector}}/\Lambda^2=1\TeV^{-2}$, is also listed for reference and includes contributions from both top quark production and decay modes.}
\newcolumntype{x}{D{,}{\,\pm\,}{4.8}}
\begin{scotch}{cxx}
Process & \multicolumn{1}{c}{$m(\Pe\PGm)<150\GeV$} & \multicolumn{1}{c}{$m(\Pe\PGm)>150\GeV$} \\
\hline
Nonprompt & 351,92 & 146,38\\
$\PW\PZ$ & 275,64 & 145,35\\
$\PZ\PZ$ & 33.2,6.5 & 13.1,2.6\\
$\PV\PV\PV$ & 17.0,8.5 & 12.0,6.0\\
$\PQt\PAQt\PW$ & 47.6,15.9 & 40.0,13.8\\
$\PQt\PAQt\PZ$ & 39.1,7.9 & 25.8,5.4\\
$\PQt\PAQt\PH$ & 28.2,4.5 & 10.0,1.6\\
$\PQt\PZ\PQq$ & 5.5,1.1 & 2.5,0.5\\
Other & 7.3,3.7 & 4.5,2.3\\[\cmsTabSkip]
\multicolumn{1}{l}{Total expected} & 805,124 & 398,58\\[\cmsTabSkip]
\multicolumn{1}{l}{Data} & \multicolumn{1}{c}{783} & \multicolumn{1}{c}{378}\\
\hline\\[-\cmsTabSkip]
CLFV & 207,15 & 4440,215\\
\end{scotch}
\label{tab:eventcount}
\end{table}

\section{Signal extraction}
\label{sec:BDT}

To separate a possible LFV signal from the background contributions, BDTs are implemented using the \textsc{XGBoost} package~\cite{Chen:2016:XST:2939672.2939785}. 
The kinematics of signal events from the top quark decay and production CLFV processes differ significantly. The most prominent difference is the presence of high-\pt leptons for the top quark production signal. Separate BDTs are therefore trained for the two SRs, top quark decay enriched ($m({\Pe\PGm})<150\GeV$) and top quark production enriched ($m({\Pe\PGm})>150\GeV$), to optimally target each signal process. 

The signal samples used for the BDT training are simulated events that correspond to the CLFV top quark production and decay signal processes. The Lorentz structure and the flavor of the up-type quark involved in the LFV interaction ($\PQu$ or $\PQc$) were shown to have a negligible impact on the kinematics of the final-state particles. Therefore, they are combined in the training. The background samples, also obtained from simulation, include prompt backgrounds from all the sources described in Section~\ref{sec:samples} and two major nonprompt backgrounds (DY and $\ttbar$ production). 

The most important input variables to the BDT model targeting the top quark decay signal are the invariant mass of the OSSF lepton pair, the number of $\PQb$-tagged jets, and the invariant mass of the flavor-violating top quark candidate. If no OSSF lepton pair exists in an event, a mass value of 0 is assigned for the purpose of BDT training. For the BDT model targeting the top quark production signal, the invariant mass of the LFV $\Pe\PGm$ pair, the \pt of the LFV electron, and the \pt of the LFV muon are the most important. 

\begin{figure*}[tbhp]
 \centering
  \includegraphics[width=0.48\textwidth]{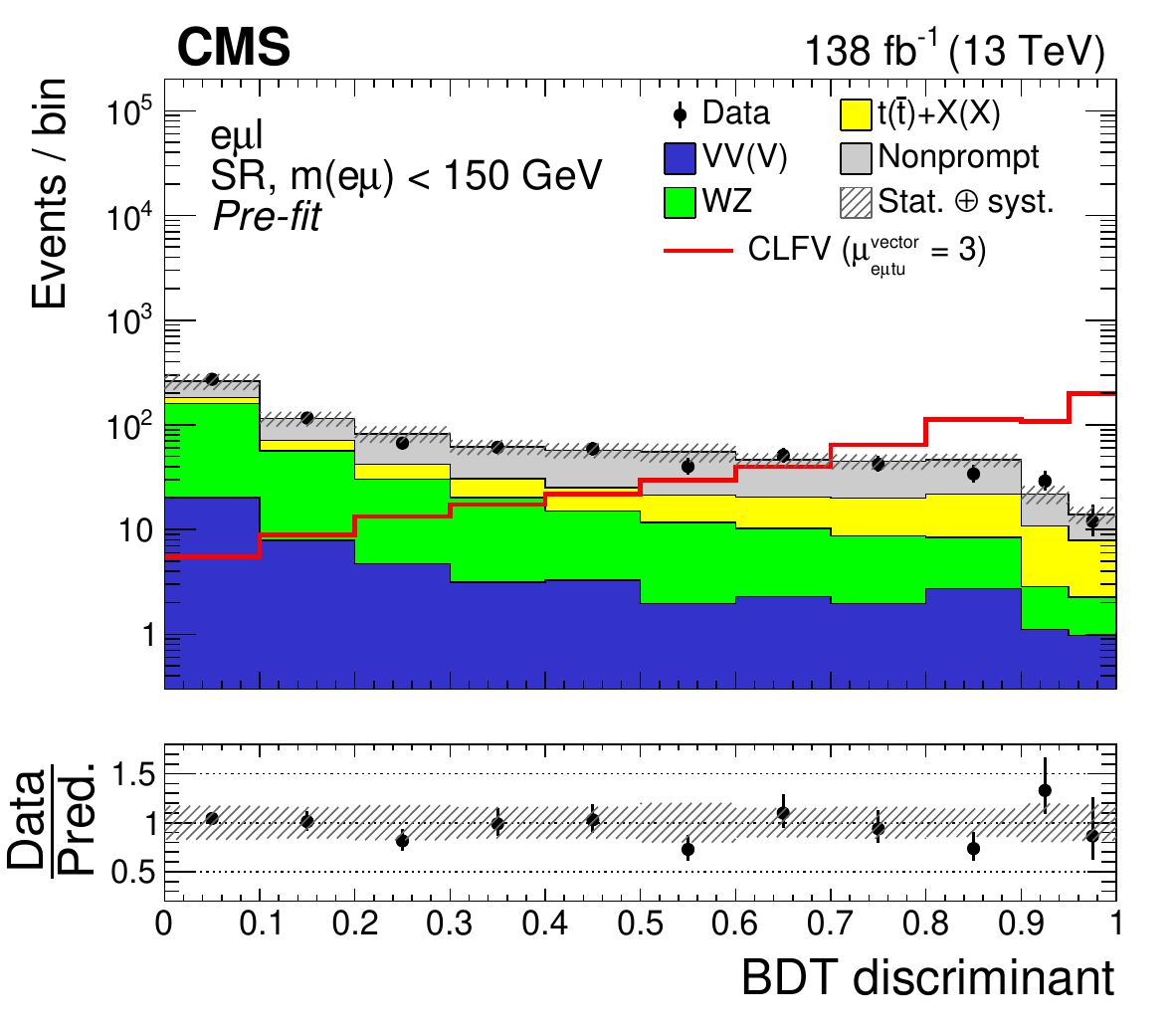}
  \includegraphics[width=0.48\textwidth]{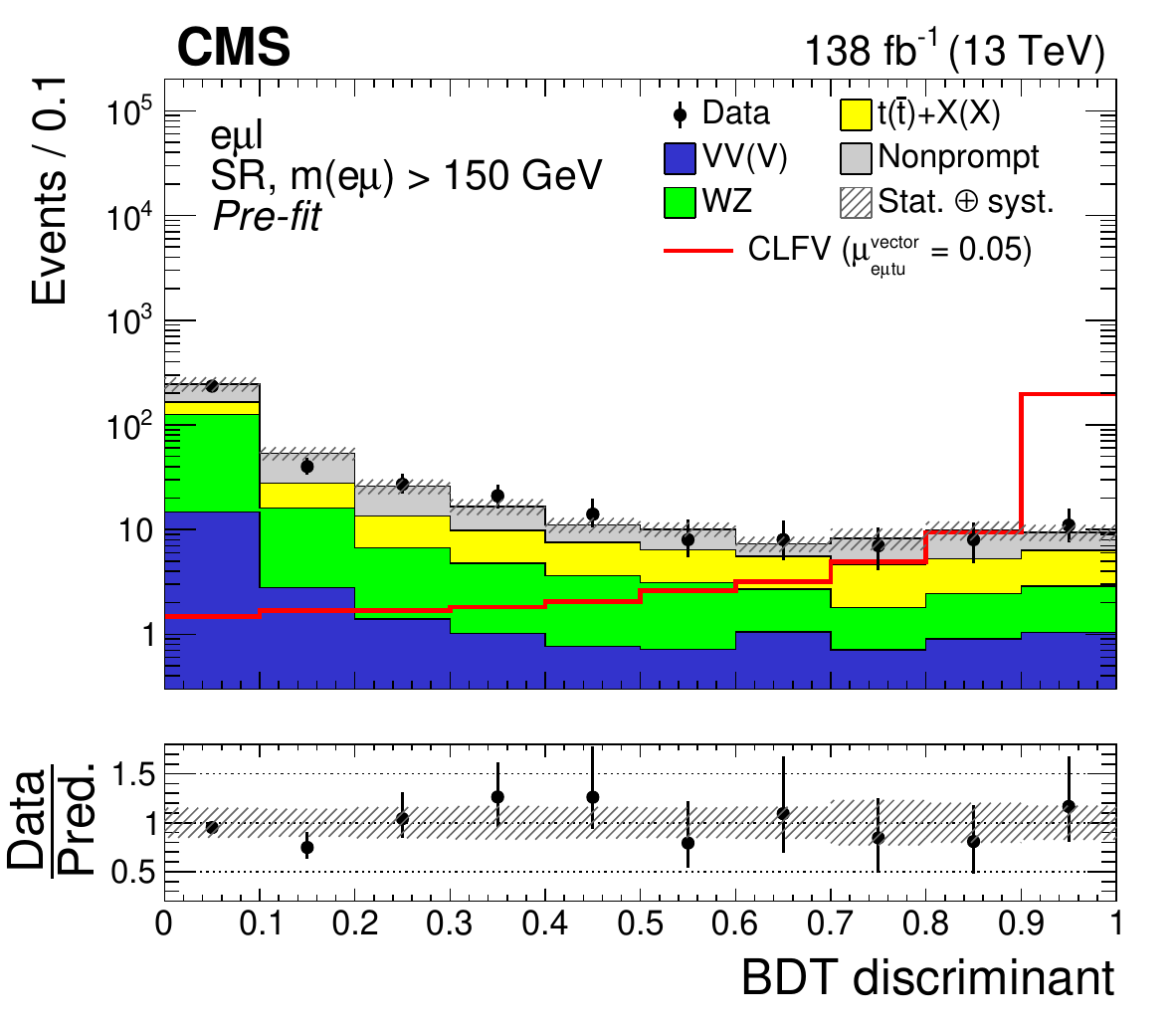}
 \caption{Distributions of the BDT discriminant targeting the CLFV top quark decay (left) and production (right) signal. Contributions from the two signal modes (production and decay) are combined within each SR and are shown as the solid red line. The pre-fit signal strength ($\mu_{\Pe\PGm\PQt\PQu}^{\text{vector}}=1$), corresponding to $C_{\Pe\PGm\PQt\PQu}^{\text{vector}}/\Lambda^2=1\TeV^{-2}$, is scaled up (down) by a factor of 3 (20) for the CLFV top quark decay (production) signal for better visualization. The hatched bands indicate statistical and systematic uncertainties in the background predictions.}
 \label{fig:bdt_output}
\end{figure*}

The BDT models targeting the two signals are used to evaluate each event in the corresponding SR. The resulting output BDT distributions are shown in Fig.~\ref{fig:bdt_output}.
The BDT discriminants efficiently distinguish signal from background events, with good agreement observed between the data and background prediction. 

\section{Systematic uncertainties}
\label{sec:systematics}

Different sources of systematic uncertainty contribute to the estimation of background events and modeling of the signal. A summary of systematic uncertainties and their average impact on predicted yields in the SRs can be found in Table~\ref{tab:repsys}. Individual uncertainties are discussed in detail below.

Uncertainties of 1.2, 2.3, and 2.5\% are assigned to the integrated luminosity for 2016, 2017, and 2018, respectively~\cite{CMS-LUM-17-003,CMS-LUM-17-004,CMS-LUM-18-002}. The uncertainties associated with the number of pileup interactions are evaluated by varying the inelastic $\Pp\Pp$ cross section by $\pm$4.6\%~\cite{Sirunyan:2018nqx}. 

The reconstruction~\cite{CMS:2020uim,CMS:2018rym}, identification, and isolation efficiencies~\cite{CMS:2021ugv} of electrons and muons are measured by applying the ``tag-and-probe'' method~\cite{CMS:2010svw} in data samples enriched in $\PZ \to \Pell\Pell$ events. Scale factors are applied to simulated events to account for the differences in efficiency between data and simulation. This correction is varied independently for electrons and muons within their corresponding uncertainties, which amount to approximately 1\% in the overall prediction. An additional uncertainty is assigned to the leptons with $\pt > 200\GeV$, as such high-\pt leptons are not well represented in these calibration studies. This uncertainty is approximately 3\% or less, affecting primarily the signal processes.  

The trigger efficiencies are measured in separate data control samples and are close to 100\% for both data and simulation due to the use of single-, double-, and triple-lepton triggers. Therefore, no correction to the trigger efficiency is applied (\ie, a data to simulation scale factor of 1 is applied to simulated events). A flat uncertainty in the scale factor of 2\% is used to cover the statistical fluctuations of the efficiency measurement. 

During the 2016--2017 data taking, a gradual shift in the timing of the inputs of the ECAL L1 trigger in the region $\abs{\eta} > 2.0$ (L1 prefiring) caused a specific inefficiency~\cite{CMS:2020cmk}. Correction factors were computed from data and applied to the acceptance evaluated by simulation. An uncertainty of 20\% is assigned to this correction.

The uncertainties due to the jet energy scale (JES) and jet energy resolution (JER) are evaluated by varying, in simulated events, the jet energies up and down by the corresponding uncertainties, as described in Ref.~\cite{CMS:2016lmd}. The uncertainties are split into multiple sources that are either correlated or uncorrelated across the years. For each of the uncertainty sources, kinematic quantities are recalculated and the event selections are reapplied. The effects of JES and JER are also propagated to the \ptmiss and $\PQb$ tagging discriminant distributions. These uncertainties amount to 1\% or less in the overall prediction.

The $\PQb$ tagging efficiencies for \PQb and light jets are measured in $\ttbar$ and $\PZ$+jets events, respectively~\cite{BTV-16-002}. The $\PQb$ tagging efficiencies in the simulation are iteratively fit to match the efficiencies in data. Multiple sources of systematic uncertainties are considered to account for the contamination of the efficiency measurement as well as statistical fluctuations. These uncertainties amount to 1\% or less on the overall prediction.

Some discrepancy was observed between data and simulation in the jet modeling VR. This is due to the underestimation  of additional jets in simulated $\PW\PZ$ and $\PZ\PZ$ samples, which has been observed in previous analyses~\cite{CMS:2018sgc,CMS:2020mpn}. An uncertainty, referred to as ``jet modeling'', is therefore derived from the jet modeling VR to cover the effect of the mismodeling. This uncertainty is estimated to be 10--20\% and affects the $\PW\PZ$ and $\PZ\PZ$ background estimates. 

The uncertainty in the nonprompt background estimate includes uncertainties in the measurement of the rates $r$ and $f$ as well as a normalization uncertainty. The uncertainty in $r$ and $f$ is estimated to be 10--30\%. 
This uncertainty includes three components: the effect of potential differences in sample composition between the measurement and application regions (this dominates and corresponds to an uncertainty of up to 20\% on $f$), accounting for prompt-lepton background contamination in the measurement region, and a statistical uncertainty.
Additionally, an uncertainty of 50\% is assigned to the leading lepton $f$ if it fails the tight requirement. This uncertainty arises from the leading lepton being matched with isolated trigger objects, which means a rather stringent isolation requirement is indirectly used for the selected leading leptons---inconsistent with the principle that leptons forming the ARs are selected with loose selection criteria. Lastly, an additional normalization uncertainty of 10\% is applied to the output of the matrix method to cover the possible variation in the normalization of the nonprompt backgrounds.

The predicted yields for all prompt backgrounds are normalized to their theoretical cross sections, which are calculated at LO for $\PQt\PW\PZ$, $\PQt\PH\PW$, and $\PQt\PH\PQq$ processes. For other processes, the theoretical cross sections calculated at NLO are used. Normalization uncertainties are assigned to all prompt backgrounds to account for uncertainties in the theoretical cross sections. A 6\% uncertainty~\cite{Campbell:2011bn} is assigned to $\PW\PZ$ and $\PZ\PZ$ processes, while for $\ttbar\PZ$ and $\ttbar\PH$, an uncertainty of 15\%~\cite{Frederix:2021agh,Kulesza:2020nfh} is used. Conservative normalization uncertainties of 20 and 30\% are assigned to the $\PQt\PZ\PQq$ and $\ttbar\PW$ processes, respectively. For other smaller background contributions, an uncertainty of 50\% is estimated.

Two theory-related systematic uncertainties (QCD scale and PDF uncertainties) are considered for the signal processes and for irreducible background events, which arise from $\PW\PZ$, $\ttbar\PV$, and $\ttbar\PH$ production. For uncertainties related to the PDFs, the procedure described in Ref.~\cite{Butterworth:2015oua} is followed: the NNPDF PDF sets~\cite{NNPDF:2014otw, NNPDF:2017mvq} are used and the root-mean-square value of the variations is taken as the PDF uncertainty. The renormalization ($\mu_{\text{R}}$) and factorization ($\mu_{\text{F}}$) scales are varied up and down by a factor of 2 individually and simultaneously, and the envelope of the variations is taken as the QCD scale uncertainty. For the QCD scale and PDF uncertainties, only the impact on the event yields through changes in the acceptance are considered since uncertainties in the overall normalization are already accounted for through uncertainties in the theoretical cross sections discussed above.

Uncertainties related to the parton shower are only considered for signal events, and only the impact on the acceptance is considered. It is evaluated by varying the $\mu_{\text{R}}$ scale up and down by a factor of 2 for the simulation of initial- and final-state radiation.

\begin{table*}[tbh]
\centering
\topcaption{Summary of systematic uncertainties and the average change in signal and overall background yields in the SRs. Uncertainties that only contain normalization effects, such as luminosity uncertainties and uncertainties in theoretical cross sections, are not included in this table.}
\begin{scotch}{lcccc}
\multirow{2}{*}{Systematic uncertainty} & \multicolumn{2}{c}{$m(\Pe\PGm)<150$\GeV} & \multicolumn{2}{c}{$m(\Pe\PGm)>150$\GeV} \\
             & Background & Signal     & Background & Signal \\
\hline
Pileup & $<$0.1\% & 0.4\% & $<$0.1\% & 0.3\%\\
Lepton reconstruction & $<$0.1\% & 0.6\% & $<$0.1\% & 1.7\%\\
Lepton identification and isolation & 0.5\% & 1.4\% & 0.5\% & 1.3\%\\
High-\pt lepton & $<$0.1\% & 0.2\% & $<$0.1\% & 3.4\%\\
Muon momentum scale and resolution & $<$0.1\% & 0.3\% & $<$0.1\% & 0.1\%\\
L1 prefiring & $<$0.1\% & 0.4\% & $<$0.1\% & 0.4\%\\
Jet energy scale and resolution& $<$0.1\% & 1.0\% & 0.6\% & 0.4\%\\
b tagging & $<$0.1\% & 0.9\% & 0.5\% & 0.5\%\\
Jet modeling & 6.5\% & \NA & 6.9\% & \NA \\
Nonprompt & 10.5\% & \NA & 8.7\% & \NA \\
PDF & $<$0.1\% & 2.3\% & $<$0.1\% & 1.3\%\\
QCD scale & 4.2\% & 2.8\% & 5.3\% & 1.4\%\\
Initial- and final-state radiation & \NA & 7.6\% & \NA & 1.0\%\\
\end{scotch}
\label{tab:repsys}
\end{table*}

\section{Results}
\label{sec:results}

A binned likelihood function $\mathcal{L}(\mu, \theta)$ is constructed to perform the statistical analysis using the binned BDT discriminant distributions. The choice of intervals for the bins in the likelihood function is the same as in Fig.~\ref{fig:bdt_output}. Six Wilson coefficients ($C_{\Pe\PGm\PQt\PQu}^{\text{tensor}}$, $C_{\Pe\PGm\PQt\PQu}^{\text{vector}}$, $C_{\Pe\PGm\PQt\PQu}^{\text{scalar}}$, $C_{\Pe\PGm\PQt\PQc}^{\text{tensor}}$, $C_{\Pe\PGm\PQt\PQc}^{\text{vector}}$, and $C_{\Pe\PGm\PQt\PQc}^{\text{scalar}}$) are considered separately in the statistical analysis, where only one of them is activated while the others are set to zero. The top quark production and decay signal modes that correspond to the same Wilson coefficient are combined. The signal strength $\mu$, defined previously in Eq.~(\ref{eq:signal_strength}), governs the cross section of the two signal modes simultaneously. 

All systematic uncertainties are incorporated into the likelihood function as nuisance parameters, denoted by $\theta$. The uncertainties that affect the shape of the BDT discriminant distributions utilize Gaussian distributions while other uncertainties that only affect the normalizations utilize log-normal distributions. The ``Barlow-Beeston lite" method \cite{Barlow:1993dm} is used to incorporate the statistical uncertainties in the signal and background predictions. 

\begin{figure*}[tbhp]
 \centering
  \includegraphics[width=0.48\textwidth]{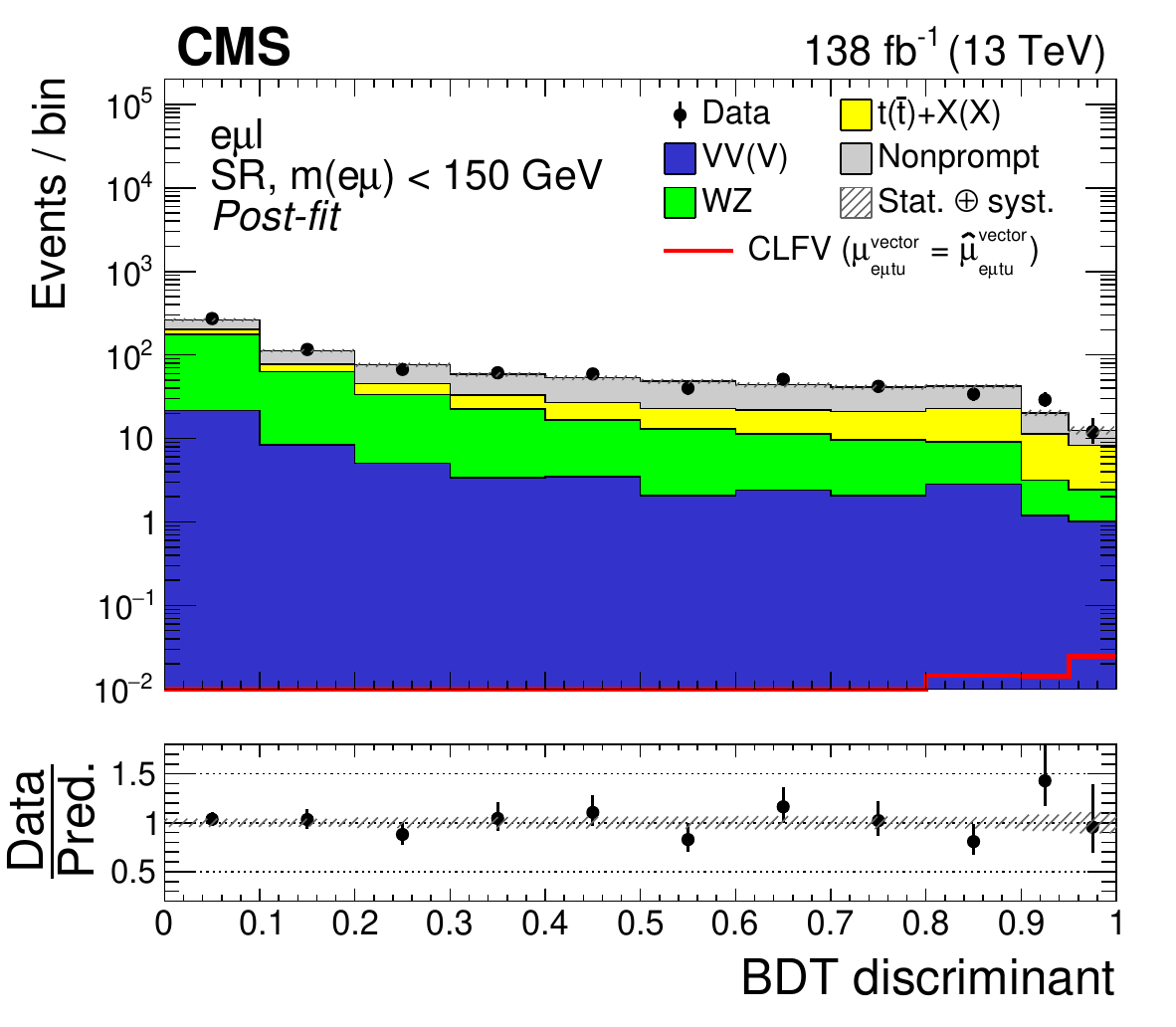}
  \includegraphics[width=0.48\textwidth]{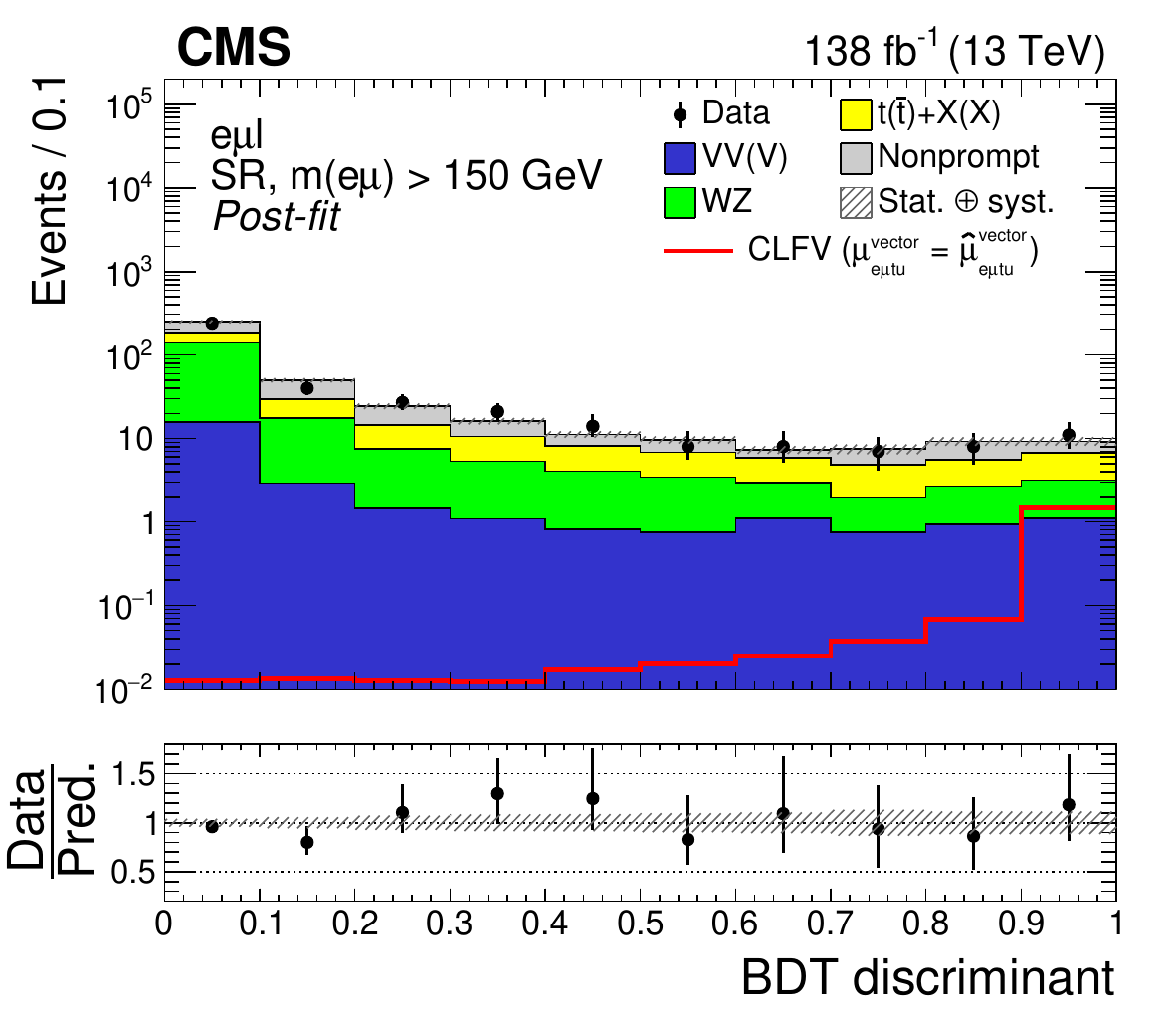}
 \caption{Distributions of the post-fit BDT discriminant targeting the CLFV top quark decay (left) and production (right) signal. Contributions from the two signal modes (production and decay) are combined within each SR and are shown as the solid red line. The post-fit signal strength ($\mu_{\Pe\PGm\PQt\PQu}^{\text{vector}}=\hat{\mu}_{\Pe\PGm\PQt\PQu}^{\text{vector}}$) is used to normalize the signal cross sections. The hatched bands indicate post-fit uncertainties (statistical and systematic) for the SM background predictions.}
 \label{fig:bdt_postfit}
\end{figure*}

A profile likelihood fit is performed simultaneously in six regions (three data-taking years and two SRs) by maximizing the likelihood function $\mathcal{L}(\mu, \hat{\theta}_{\mu})$, where $\hat{\theta}_{\mu}$ are the values of the nuisance parameters that maximize the likelihood for a specific signal strength. The post-fit distributions of the BDT discriminants are shown in Fig.~\ref{fig:bdt_postfit}. The limited number of events in the signal region is the dominant uncertainty in this analysis. Of the post-fit systematic uncertainties, the largest impact is from the statistical uncertainty associated with the limited number of events in the ARs constructed for the nonprompt background estimate. Other important systematic uncertainties include the normalization uncertainties associated with the VV and $\ttbar \PW$ backgrounds and the QCD scale uncertainty used for the $\ttbar \PW$ background.

The results are interpreted as 95\% confidence level (\CL) upper limits on the signal strength $\mu$, which is equivalent to $(C/\Lambda^2)^2$, as shown in Eq.~(\ref{eq:signal_strength}). The limit setting procedure uses a modified frequentist $\CLs$ method~\cite{Junk:1999kv,Read2002} as the criterion, which is based on a profile likelihood ratio test statistic. An asymptotic approximation~\cite{Cowan:2010js} is used to determine the distributions of the test statistic. By convention, positive Wilson coefficients are assumed, and the one-dimensional upper limits on a given Wilson coefficient, $C_a/\Lambda^2$, are obtained by taking the square root of the upper limits on the corresponding signal strength $\mu_a$. The branching fractions, $\mathcal{B} (\PQt \to \Pe\PGm\PQq)$ with $\PQq=\PQu$ or $\PQc$, are obtained using the equation below, taken from Ref.~\cite{Kile_2008}:

\begin{equation}
\label{eq:BR}
\mathcal{B}(\PQt \to \Pe\PGm\PQq) = 
\left\{ 
 \begin{array}{l}
 \frac{|C_{\Pe\PGm\PQt\PQq}^{\text{tensor}}|^2}{\Lambda^4}\frac{m_{\PQt}^5}{64\pi^3\Gamma_{\PQt}^{\text{SM}}} 
\\
 \frac{|C_{\Pe\PGm\PQt\PQq}^{\text{vector}}|^2}{\Lambda^4}\frac{m_{\PQt}^5}{384\pi^3\Gamma_{\PQt}^{\text{SM}}} \\
 \frac{|C_{\Pe\PGm\PQt\PQq}^{\text{scalar}}|^2}{\Lambda^4}\frac{m_{\PQt}^5}{3072\pi^3\Gamma_{\PQt}^{\text{SM}}}
\end{array}
\right.
\end{equation}
where the top quark mass $m_{\PQt}$ and width $\Gamma_{\PQt}^{\text{SM}}$ are assumed to be 172.5\GeV and 1.33\GeV, respectively. 

\begin{figure*}[tbhp]
 \centering
  \includegraphics[width=0.48\textwidth]{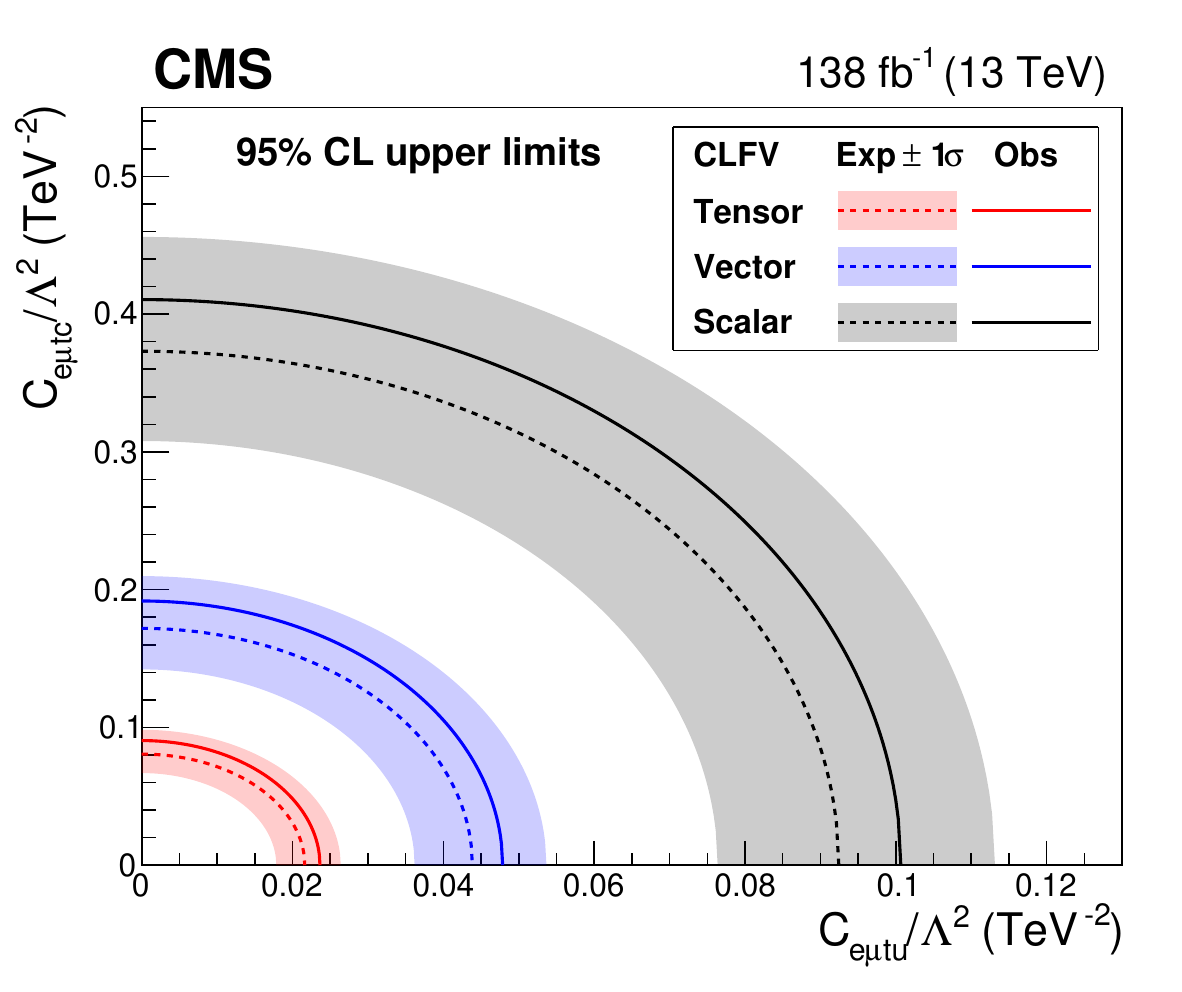}
  \includegraphics[width=0.48\textwidth]{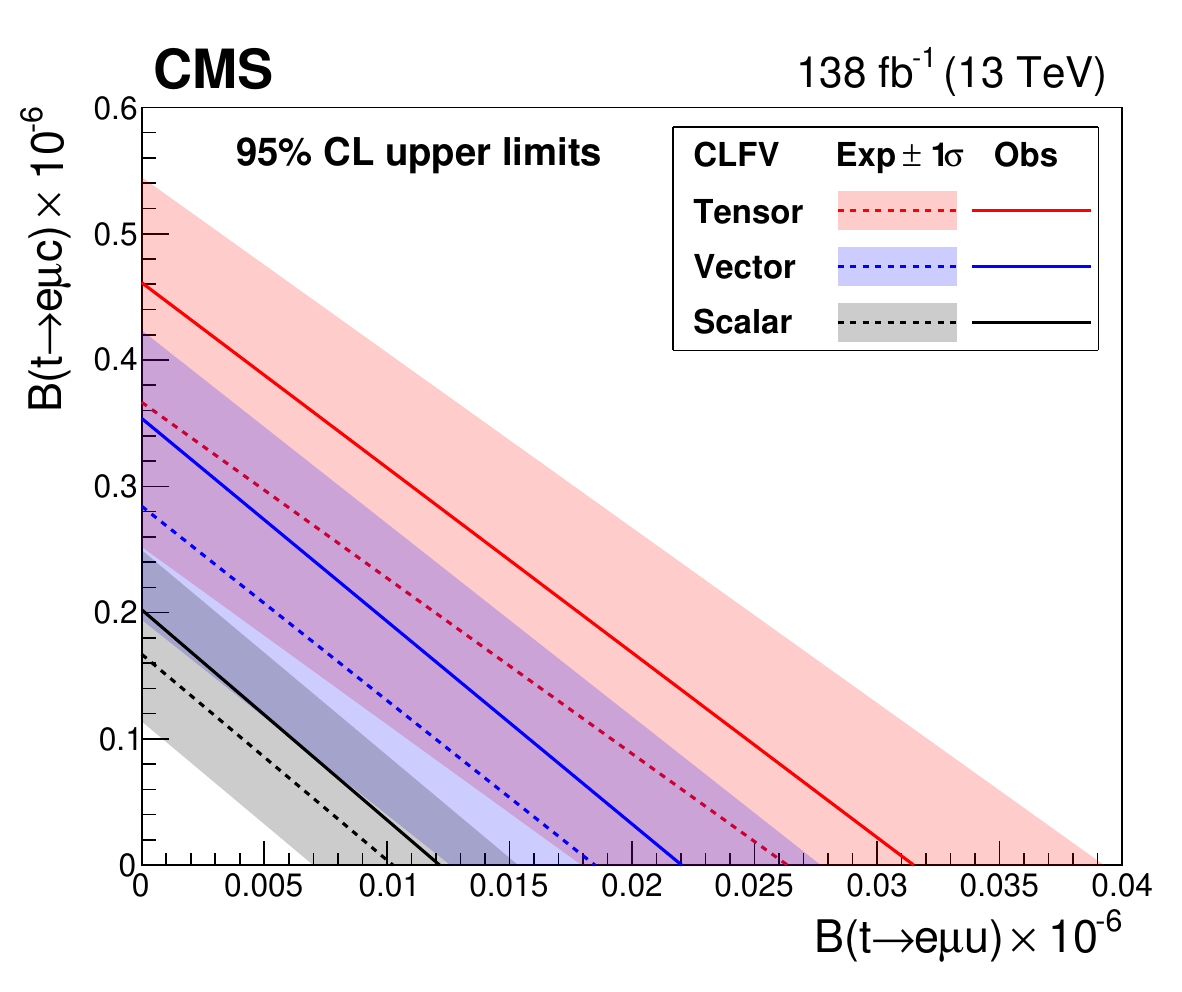}
 \caption{Interpolated 95\% \CL upper limits on the Wilson coefficients (left) and the branching fractions (right). The observed (expected) upper limits for tensor-, vector-, and scalar-like CLFV interactions are shown in red, blue, and black solid (dotted) lines, respectively. The shaded bands contain 68\% of the distribution of the expected upper limits.}
 \label{fig:2dlimit}
\end{figure*}

The resulting one-dimensional limits are summarized in Table~\ref{tab:limit}. Tabulated results are provided in the HEPData record for this analysis~\cite{hepdata}. The upper limits on Wilson coefficients associated to operators with a tensor-like Lorentz structure are more stringent because the predicted cross sections for tensor-like operators are higher when all Wilson coefficients are set to the same value (\ie, $C_{a}/\Lambda^2=1\TeV^{-2}$), as shown in Table~\ref{tab:XS}. Signal processes initiated by an up quark receive large contributions from valence quarks coming from protons, which lead to larger cross sections, and consequently limits on Wilson coefficients involving an up quark are more stringent that those with a charm quark. In addition to the cross sections, the sensitivity for the branching fractions $\mathcal{B} (\PQt \to \Pe\PGm\PQq)$ are also correlated to several constants specified in Eq.~(\ref{eq:BR}). Relative to the vector- and scalar-like interactions, the tensor-like CLFV interactions have the least stringent limits on the corresponding branching fractions due to the small combinatorial factor that appears in the denominator of Eq.~(\ref{eq:BR}). The one-dimensional limits on $\Pe\PGm\PQt\PQu$ and $\Pe\PGm\PQt\PQc$ couplings are interpolated to be visualized in two-dimensional planes, which are shown in Fig.~\ref{fig:2dlimit}. 

\begin{table*}[tbh]
\centering
\topcaption{Expected and observed upper limits at 95\% \CL on Wilson coefficients and the branching fractions for tensor-, vector-, and scalar-like CLFV interactions. The intervals that contain 68\% of the distribution of the expected upper limits are shown in parentheses.}
\begin{scotch}{cccccc}
CLFV       & Lorentz  & \multicolumn{2}{c}{$C_{\Pe\PGm\PQt\PQq}/\Lambda^2~(\TeVns^{-2})$} & \multicolumn{2}{c}{$\mathcal{B} (\PQt \to \Pe\PGm\PQq) \times 10^{-6}$} \\
coupling  & structure & Exp. (68\% \CL range) & Obs. & Exp. (68\% \CL range) & Obs. \\
\hline
\multirow{3}{*}{$\Pe\PGm\PQt\PQu$}& Tensor & 0.022 (0.018--0.026) & 0.024 & 0.027 (0.018--0.040) & 0.032\\
& Vector & 0.044 (0.036--0.054) & 0.048 & 0.019 (0.013--0.028) & 0.022\\
& Scalar & 0.093 (0.077--0.114) & 0.101 & 0.010 (0.007--0.016) & 0.012\\
\hline
\multirow{3}{*}{$\Pe\PGm\PQt\PQc$} & Tensor & 0.084 (0.069--0.102) & 0.094 & 0.396 (0.272--0.585) & 0.498\\
 & Vector & 0.175 (0.145--0.214) & 0.196 & 0.296 (0.203--0.440) & 0.369\\
 & Scalar & 0.385 (0.318--0.471) & 0.424 & 0.178 (0.122--0.266) & 0.216\\
\end{scotch}
\label{tab:limit}
\end{table*}

\section{Summary}
\label{sec:summary}

This paper presents results from a search for charged-lepton flavor violation in both top quark production and decay processes. The data used were collected by the CMS experiment during 2016--2018 and correspond to an integrated luminosity of 138\fbinv. Events were selected for analysis if they contain exactly three charged leptons---one electron and one muon of opposite electric charge as well as one additional electron or muon. Events must also contain at least one jet of which no more than one is associated with a bottom quark. An effective field theory approach is used for parametrizing the charged-lepton flavor violating interactions. Boosted decision trees are used to distinguish a possible signal from the background. No significant excess is observed over the prediction from the standard model. Upper limits at the 95\% confidence level are set on the branching fractions involving up (charm) quarks, $\PQt\to\Pe\PGm\PQu$ ($\PQt\to\Pe\PGm\PQc$), of $0.032 (0.498) \times 10^{-6}$, $0.022 (0.369) \times 10^{-6}$, and $0.012 (0.216) \times 10^{-6}$ for tensor, vector, and scalar interactions, respectively. These limits constitute the most stringent ones to date on these processes, improving the existing limits by roughly one order of magnitude.

\begin{acknowledgments}
We congratulate our colleagues in the CERN accelerator departments for the excellent performance of the LHC and thank the technical and administrative staffs at CERN and at other CMS institutes for their contributions to the success of the CMS effort. In addition, we gratefully acknowledge the computing centers and personnel of the Worldwide LHC Computing Grid and other centers for delivering so effectively the computing infrastructure essential to our analyses. Finally, we acknowledge the enduring support for the construction and operation of the LHC, the CMS detector, and the supporting computing infrastructure provided by the following funding agencies: SC (Armenia), BMBWF and FWF (Austria); FNRS and FWO (Belgium); CNPq, CAPES, FAPERJ, FAPERGS, and FAPESP (Brazil); MES and BNSF (Bulgaria); CERN; CAS, MoST, and NSFC (China); MINCIENCIAS (Colombia); MSES and CSF (Croatia); RIF (Cyprus); SENESCYT (Ecuador); MoER, ERC PUT and ERDF (Estonia); Academy of Finland, MEC, and HIP (Finland); CEA and CNRS/IN2P3 (France); SRNSF (Georgia); BMBF, DFG, and HGF (Germany); GSRI (Greece); NKFIH (Hungary); DAE and DST (India); IPM (Iran); SFI (Ireland); INFN (Italy); MSIP and NRF (Republic of Korea); MES (Latvia); LAS (Lithuania); MOE and UM (Malaysia); BUAP, CINVESTAV, CONACYT, LNS, SEP, and UASLP-FAI (Mexico); MOS (Montenegro); MBIE (New Zealand); PAEC (Pakistan); MES and NSC (Poland); FCT (Portugal);  MESTD (Serbia); MCIN/AEI and PCTI (Spain); MOSTR (Sri Lanka); Swiss Funding Agencies (Switzerland); MST (Taipei); MHESI and NSTDA (Thailand); TUBITAK and TENMAK (Turkey); NASU (Ukraine); STFC (United Kingdom); DOE and NSF (USA).

\hyphenation{Rachada-pisek} Individuals have received support from the Marie-Curie program and the European Research Council and Horizon 2020 Grant, contract Nos.\ 675440, 724704, 752730, 758316, 765710, 824093, and COST Action CA16108 (European Union); the Leventis Foundation; the Alfred P.\ Sloan Foundation; the Alexander von Humboldt Foundation; the Science Committee, project no. 22rl-037 (Armenia); the Belgian Federal Science Policy Office; the Fonds pour la Formation \`a la Recherche dans l'Industrie et dans l'Agriculture (FRIA-Belgium); the Agentschap voor Innovatie door Wetenschap en Technologie (IWT-Belgium); the F.R.S.-FNRS and FWO (Belgium) under the ``Excellence of Science -- EOS" -- be.h project n.\ 30820817; the Beijing Municipal Science \& Technology Commission, No. Z191100007219010 and Fundamental Research Funds for the Central Universities (China); the Ministry of Education, Youth and Sports (MEYS) of the Czech Republic; the Shota Rustaveli National Science Foundation, grant FR-22-985 (Georgia); the Deutsche Forschungsgemeinschaft (DFG), under Germany's Excellence Strategy -- EXC 2121 ``Quantum Universe" -- 390833306, and under project number 400140256 - GRK2497; the Hellenic Foundation for Research and Innovation (HFRI), Project Number 2288 (Greece); the Hungarian Academy of Sciences, the New National Excellence Program - \'UNKP, the NKFIH research grants K 124845, K 124850, K 128713, K 128786, K 129058, K 131991, K 133046, K 138136, K 143460, K 143477, 2020-2.2.1-ED-2021-00181, and TKP2021-NKTA-64 (Hungary); the Council of Science and Industrial Research, India; ICSC -- National Research Center for High Performance Computing, Big Data and Quantum Computing, funded by the EU NexGeneration program (Italy); the Latvian Council of Science; the Ministry of Education and Science, project no. 2022/WK/14, and the National Science Center, contracts Opus 2021/41/B/ST2/01369 and 2021/43/B/ST2/01552 (Poland); the Funda\c{c}\~ao para a Ci\^encia e a Tecnologia, grant CEECIND/01334/2018 (Portugal); the National Priorities Research Program by Qatar National Research Fund; MCIN/AEI/10.13039/501100011033, ERDF ``a way of making Europe", and the Programa Estatal de Fomento de la Investigaci{\'o}n Cient{\'i}fica y T{\'e}cnica de Excelencia Mar\'{\i}a de Maeztu, grant MDM-2017-0765 and Programa Severo Ochoa del Principado de Asturias (Spain); the Chulalongkorn Academic into Its 2nd Century Project Advancement Project, and the National Science, Research and Innovation Fund via the Program Management Unit for Human Resources \& Institutional Development, Research and Innovation, grant B37G660013 (Thailand); the Kavli Foundation; the Nvidia Corporation; the SuperMicro Corporation; the Welch Foundation, contract C-1845; and the Weston Havens Foundation (USA).
\end{acknowledgments}

\bibliography{auto_generated}\cleardoublepage \appendix\section{The CMS Collaboration \label{app:collab}}\begin{sloppypar}\hyphenpenalty=5000\widowpenalty=500\clubpenalty=5000
\cmsinstitute{Yerevan Physics Institute, Yerevan, Armenia}
{\tolerance=6000
A.~Hayrapetyan, A.~Tumasyan\cmsAuthorMark{1}\cmsorcid{0009-0000-0684-6742}
\par}
\cmsinstitute{Institut f\"{u}r Hochenergiephysik, Vienna, Austria}
{\tolerance=6000
W.~Adam\cmsorcid{0000-0001-9099-4341}, J.W.~Andrejkovic, T.~Bergauer\cmsorcid{0000-0002-5786-0293}, S.~Chatterjee\cmsorcid{0000-0003-2660-0349}, K.~Damanakis\cmsorcid{0000-0001-5389-2872}, M.~Dragicevic\cmsorcid{0000-0003-1967-6783}, A.~Escalante~Del~Valle\cmsorcid{0000-0002-9702-6359}, P.S.~Hussain\cmsorcid{0000-0002-4825-5278}, M.~Jeitler\cmsAuthorMark{2}\cmsorcid{0000-0002-5141-9560}, N.~Krammer\cmsorcid{0000-0002-0548-0985}, D.~Liko\cmsorcid{0000-0002-3380-473X}, I.~Mikulec\cmsorcid{0000-0003-0385-2746}, J.~Schieck\cmsAuthorMark{2}\cmsorcid{0000-0002-1058-8093}, R.~Sch\"{o}fbeck\cmsorcid{0000-0002-2332-8784}, D.~Schwarz\cmsorcid{0000-0002-3821-7331}, M.~Sonawane\cmsorcid{0000-0003-0510-7010}, S.~Templ\cmsorcid{0000-0003-3137-5692}, W.~Waltenberger\cmsorcid{0000-0002-6215-7228}, C.-E.~Wulz\cmsAuthorMark{2}\cmsorcid{0000-0001-9226-5812}
% ...
\par}
\vskip\cmsinstskip
\dag:~Deceased\\
$^{1}$Also at Yerevan State University, Yerevan, Armenia\\
$^{2}$Also at TU Wien, Vienna, Austria\\
% ...
\end{sloppypar}
\end{document}